\documentclass[a4paper,fleqn,usenatbib]{mnras}

\usepackage[T1]{fontenc}
\usepackage{ae,aecompl}

\usepackage{graphicx}	
\usepackage{amsmath}	
\usepackage{amssymb}	
\usepackage{pdflscape}

\newcommand{\msun}{M$_{\odot}$} 
\newcommand{\kms}{\,km\,s$^{-1}$} 
\newcommand{\ha}{H$\alpha$} 

\title[GRS 1716-249]{The orbital period, black hole mass and distance to the X-ray transient GRS 1716-249 (=N Oph 93)}
\author[J. Casares et al.]{J. Casares$^{1,2}$\thanks{E-mail: jorge.casares@iac.es}, 
 I.V. Yanes-Rizo$^{1,2}$, M.A.P. Torres$^{1,2}$, T.M.C. Abbott$^{3}$, M. Armas 
 \newauthor Padilla$^{1,2}$, P.A. Charles$^{4,5}$ , V.A. C\'uneo$^{1,2}$, T. Mu\~noz-Darias$^{1,2}$, P.G. Jonker$^{6,7}$ and 
 \newauthor K. Maguire$^{8}$
\\
$^{1}$Instituto de Astrof\'\i{}sica de Canarias, 38205 La Laguna, Tenerife, Spain\\
$^{2}$Departamento de Astrof\'isica, Universidad de La Laguna, E-38206 La Laguna, Tenerife, Spain\\
$^{3}$Cerro Tololo Inter-American Observatory, NSF's National Optical-Infrared Astronomy Research Laboratory, Casilla 603, La Serena, Chile\\
$^{4}$Department of Physics, Astrophysics, University of Oxford, Denys Wilkinson Building, Keble Road, Oxford OX1 3RH, UK\\
$^{5}$Department of Physics and Astronomy, University of Southampton, Southampton SO17 1BJ, UK\\
$^{6}$Department of Astrophysics/IMAPP, Radboud University, PO Box 9010, 6500 GL Nijmegen, The Netherlands \\
$^{7}$SRON, Netherlands Institute for Space Research, Sorbonnelaan 2, 3584 CA, Utrecht, The Netherlands \\
$^{8}$School of Physics, Trinity College Dublin, The University of Dublin, Dublin 2, D02 PN40 Ireland
}

\date{Accepted XXX. Received YYY; in original form ZZZ}

\pubyear{2023}

\begin{document}
\label{firstpage}
\pagerange{\pageref{firstpage}--\pageref{lastpage}}
\maketitle

\begin{abstract}
We present evidence for a 0.278(8) d (=6.7 h) orbital period in the X-ray transient GRS 1716-249 (=N Oph 93),   
based on a superhump modulation detected during the 1995 mini-outburst plus ellipsoidal variability in quiescence. 
With a quiescent magnitude of $r=23.19\pm0.15$ N Oph 93 is too faint to warrant a full dynamical study through 
dedicated time-resolved spectroscopy. Instead, we apply the FWHM-$K_2$ correlation to the disc 
\ha~emission line detected in Gran Telescopio Canarias spectra and obtain $K_2=521\pm52$ km s$^{-1}$. 
This leads to a mass function $f(M)=4.1\pm1.2$ \msun, thus indicating the presence of a black hole in this historic 
X-ray transient. Furthermore, from the depth of the \ha~trough and the quiescent light curve  
we constrain the binary inclination to $i=61\pm15^{\circ}$, while the detection of superhumps sets an upper limit 
to the donor to compact star mass ratio $q=M_{2}/M_{1}\lesssim0.25$.  Our de-reddened $(r-i)$ colour is consistent 
with a $\approx$K6 main sequence star that fills its Roche lobe in a  0.278 d orbit. 
Using all this information we derive a compact object mass $M_{1}=6.4^{+3.2}_{-2.0}$ \msun~at 68 per cent confidence. 
We also constrain the distance to GRS 1716-249 to 6.9$\pm$1.1 kpc, placing the binary $\sim$0.8 kpc 
above the Galactic Plane, in support of a large natal kick. 
\end{abstract}




\section{Introduction}
\label{intro}

The Milky Way contains an estimated population of $\approx10^{3}-10^{4}$ stellar-mass black holes (BH) 
accreting from a late-type companion star \citep{yungelson06,kiel06}. 
As a result of their low average mass-accretion rates these BH binaries are X-ray transients 
(XRTs) and are routinely discovered at a rate of $\sim1.7$ yr$^{-1}$ during month-long episodes of increased 
accretion activity \citep{mcclintock06, belloni11, corral16}. In the interim, XRTs  
hibernate for decades in a quiescent state, with typical luminosities L$_{\rm X}\lesssim10^{32}$ erg s$^{-1}$. 
Optical/near-infrared studies performed during quiescence typically reveal the spectrum of the companion star, which is 
used as a probe to weigh the accreting star (see \citealt{casares-jonker14} for a review of the techniques and inferred BH masses). 

The latest version of the BlackCAT catalogue (https://www.astro.puc.cl/BlackCAT/index.php; see \citealt{corral16}) 
contains 70 BH XRTs discovered since the start of X-ray missions, with 19 BHs dynamically confirmed 
(i.e. with dynamical masses in excess of 3 \msun, the upper bound of a non-rotating neutron star). 
In addition, by using  a scaling relation between the full-width-at-half-maximum (FWHM) of the \ha~line 
(formed in the accretion disc) and the radial velocity semi-amplitude of the companion star $K_2$ (see \citealt{casares15}) 
indirect evidence for dynamical  
BHs has been reported in four other XRTs: Swift J1357.2-0933 \citep{mata15}, KY TrA \citep{zurita15}, Swift J1753.5-0127 
\citep{shaw16} and. MAXI J1659-152 \citep{torres21}. 
In all these cases the  donor stars are undetected because the 
optical counterparts are either too faint ($r\gtrsim23$) or strongly contaminated by the light of the accretion disc or a 
close interloper. 
Other properties of the \ha~line can also be exploited to obtain reliable 
estimates of the donor to compact star mass ratio $q=M_{2}/M_{1}$ and the binary inclination \citep{casares16,casares22}. 
In the current paper, we apply this same strategy  to the historical XRT GRS 1716-249 
(=N Oph 93), which has a challenging quiescent magnitude $r$=23.2. 

GRS 1716-249 (GRS1716 hereafter) was discovered in September 1993 during an X-ray outburst with maximum 
flux  $F_{\rm X}$ (20-100 keV) $\sim$1.4 Crab \citep{ballet93,harmon93}. It was promptly 
classified as a strong BH candidate on the basis of its X-ray and radio properties, reminiscent of 
those exhibited by dynamical BHs in the hard state. In particular, GRS1716 displayed a hard power-law spectrum, strong 
variability at short time-scales, the presence of a low-frequency type-C 
quasi-periodic oscillation (QPO), a flat radio spectrum and bright radio flares, 
correlated with X-rays   \citep{dellavalle94,vanderhooft96,hjellming96,revnivtsev98}.
The optical counterpart (N Oph 93, V2293 Oph) was identified with a V$\sim16.6$ star 
two days after the outburst peak. No quiescent counterpart brighter than $V\sim21$ could, however, be detected in 
pre-outburst UK and European Southern Observatory (ESO) Schmidt plates \citep{dellavalle94}. 
A tentative $\sim$14.7 h superhump\footnote{Optical modulation caused by the tidal stress of the companion star on an eccentric 
accretion disc \citep{whitehurst91}.}  period was proposed from sparse photometry collected in 
$\sim$10-60 min $V$-band nightly  blocks over ten consecutive nights during outburst (\citealt{masetti96}, M96 hereafter).  
The same authors proposed a BH mass $\geq4.9$ \msun, while \cite{dellavalle94} argue for a distance of 2.4$\pm$0.4 kpc 
after comparing the peak optical brightness with other BH XRTs. 
Furthermore, GRS1716 exhibited renewed activity in 1994 and 1995 at 10--20 per cent level of the 1993 main outburst, before 
settling down in a quiescent state  \citep{churazov94,borozdin94,harmon94,borozdin95}. 

After dwelling for 23 years in quiescence, 
GRS1716 was found to be active again in December 2016 \citep{masumitsu16,negoro16}. {\it Swift} and {\it NuSTAR} 
spectral analysis of the new outburst showed this time evidence for a weak thermal disc component. A horizontal 
displacement across the hardness-intensity diagram towards softer colours was also witnessed, although 
failing to make a full transition to the canonical soft state \citep{bassi19,jiang20}. 
Type-C QPOs were detected once more in the hard state \citep{bharali19}, a classic signature of BH XRTs.  
Meanwhile, simultaneous  radio observations found the system 
in the radio-quiet branch of the radio/X-ray 
luminosity plane through the entire outburst, following L$_{\rm R}\propto{\rm L_{X}}^{1.4}$  \citep{bassi19}. 
On the other hand, optical spectroscopy obtained over the course of the outburst decay provided evidence for 
conspicuous disc outflows and a possible inflow signature, perhaps caused by a failed wind \citep{cuneo20}. 
Finally, in September 2017 GRS1716 returned to quiescence, where it has remained ever since \citep{bassi19}. 

Recently, \cite{saikia22} reported a detection of the quiescent counterpart of N. Oph 93 with an $i'$-band magnitude of 
21.39$\pm$0.15. Based on the global optical and X-ray properties observed during the 2016-2017 outburst they also 
challenge the 2.4 kpc distance of \cite{dellavalle94} and propose a new 
value  in the range $\sim$4--8 kpc.  Here in this paper we present 
photometric evidence of a 6.7 h orbital period, based on superhump light curves, recovered from the 1995 mini-outburst,  
plus an ellipsoidal modulation obtained through deep $i$-band photometry in 2021. 
In addition, we  report on the presence of an \ha~emission line in quiescent spectroscopy 
and use it to derive dynamical constraints on GRS1716.  Our photometric and spectroscopic data 
strongly advocate for the presence of a BH in GRS1716, and lend support to a binary distance of $\sim$6.9 kpc, 
in good agreement with the results of \cite{saikia22}.  

\section{Observations and data reduction}
\label{sec:obs}
 
\subsection{Photometry}
\label{sec:phot}

In an attempt to measure the orbital period of GRS1716, 
we first rescued $r$-band photometry obtained by us with the Danish 1.54-m telescope at the La Silla Observatory in Chile 
during the 1995 mini-outburst. Observations were taken with the Tektronix TK1024M CCD on the nights of May 7--9 1995. 
A total of 147$\times$300 s images were collected with the Gunn $R$-band filter over the three nights, with an average   
coverage of 5 h per night. Nights were clear, with a median seeing varying between 0.8 and 1.2 arcsec. Differential 
photometry was performed relative to the local comparison star Pan-STARRS ID 77972599165136381, which has 
similar brightness to the target. Unfortunately, the reduced images are lost and the raw data can no longer be retrieved 
since ESO does not keep an archive for the Danish telescope. However, we have managed to recover the data points 
after digitizing our printed plots of the three nightly light curves. Based on the magnitude of the comparison star quoted by 
Pan-STARRS \citep{tonry12} we estimate that GRS176 had a mean R-Johnson magnitude of 17.05$\pm$0.02 at the time.    

Photometric observations of GRS1716 were also programmed during the quiescent phase in 2019--2021.  
First, we obtained exploratory images using the Auxiliary-port CAMera (ACAM) on the 4.2 m William Herschel 
Telescope (WHT) at the Roque de los Muchachos Observatory in La Palma, on the nights of July 24 and Aug 23 2019. 
ACAM has a circular field of view with 8.3 arcmin diameter and it is equipped with a low-fringing 2k$\times$4k EEV CCD 
that gives a platescale of 0.25 arcsec pix$^{-1}$.  One 600 s integration in Sloan $r$ and another one in  Sloan $i$ were 
obtained  on July 24 and three 300 s $i$-band images on Aug 23under photometric conditions, with seeing $\simeq$0.9 arcsec. 

Subsequently, an intensive photometric campaign was performed   
with the Goodman High Throughput Spectrograph in imaging mode on the Southern Astrophysical Research (SOAR) 
4.1m Telescope at Cerro Pach\'on in Chile. A total of 82$\times$600 s integrations were obtained 
with the Sloan-$i$ filter on the nights of May 31 and June 1  2021, covering $\sim$9 h per night. One Sloan-$r$ image was  
also collected each night to measure the colour of GRS 1716. We employed the red camera of the instrument, 
a 4k$\times$4k e2v 231-84 CCD which provides a circular field of view of 7.2 arcmin diameter. 
We used 2$\times$2 binning which results in a platescale of 0.3 arcsec pix$^{-1}$.  
The SOAR observations had to be interrupted for about 100 min every  
night due to telescope pointing limitations at altitude $>88^{\circ}$ caused by its Alt-Az design.  
The first frames of each night were affected by problems with telescope tracking and the active optics 
system.  These were rejected as they proved to be unusable for aperture or point spread 
function (PSF) photometry because of the poor image quality and crowded nature of the field close to the target. 
This leaves a total of 44 useful images for the photometric analysis with 0.9--1.2 arcsec seeing. 
A log of the photometric observations is presented in Table~\ref{tab:tab1}.  

 \begin{table}
	\centering
	\caption{Journal of photometric observations}
	\label{tab:tab1}
	\begin{tabular}{lccc}
		\hline
Date & Instrument& $r$ & $i$  \\ 
 		\hline
 2019 July 24 & WHT+ACAM  & 600 s & 600 s \\
 2019 Aug 23 &              ,,        & - & 3$\times$300 s \\
		\hline
 2021 May 31  & SOAR+Goodman & 900 s & 37$\times$600 s \\
  2021 June 1  &              ,,        &  600 s & 45$\times$600 s \\
                 \hline
	\end{tabular}
\end{table}

The ACAM and SOAR images were reduced in the standard way, with bias subtraction, flat-field correction and stellar alignment using 
{\small{{\texttt{IRAF}}}}\footnote{{\small{{\texttt{IRAF}}}}  is distributed by the National Optical Astronomy Observatories, 
which are operated by AURA, Inc., under cooperative agreement with the National Science Foundation.} tasks. 
For the flat-field correction we used a set of sky images obtained during evening twilight. 
We extracted the stellar counts using PSF photometry with DAOPHOT \citep{stetson87}. 
More details of this process are given in Sect.~\ref{sec:2021}.

\subsection{Spectroscopy}
\label{sec:spect}

We also collected spectroscopic observations of GRS1716 
on the night of June 22 2020 using the Optical System 
for Imaging and low-Intermediate-Resolution Integrated Spectroscopy (OSIRIS; \citealt{cepa00}) at the 10.4m Gran Telescopio 
Canarias (GTC). These observations were obtained as part of the DDT Director's Discretionary Time program GTC2020-153. 
Four 1800 s spectra were collected using grism R1000R and a 1.0 arcsec slit, resulting in a wavelength coverage 
of 5000--10000 \AA~at 8.3 \AA~resolution (378 \kms~at \ha). 
The slit was oriented at parallactic angle to minimize light losses across the spectrum caused by  
atmospheric refraction. Seeing conditions were $\leq$0.9 arcsec through the course of the integrations.  
The 1-D spectra were extracted using optimal 
extraction routines within PAMELA \citep{marsh89} after standard debiasing and flat-field corrections. 
Observations of HgAr+Ne lamps were employed to derive a pixel-to-wavelength calibration through a 4th order polynomial 
fit to 37 lines across the entire wavelength range. Small flexure corrections, obtained from the position of 
the \ion{O}{I} 5577.34 and 6300.30 \AA~sky lines, were applied to individual spectra.  
The resulting spectra were imported to {\small{{\texttt{MOLLY}}}}, resampled to a uniform velocity scale of 104 km s$^{-1}$ and 
shifted into the heliocentric velocity frame. 
Each spectrum was also rectified through division by a 3rd order spline fit to the continuum level. 

\section{Analysis and Results}
\label{sec:analysis}

 \subsection{A superhump modulation in 1995}
\label{sec:superhump} 

\begin{figure}
	\includegraphics[angle=0,width=\columnwidth]{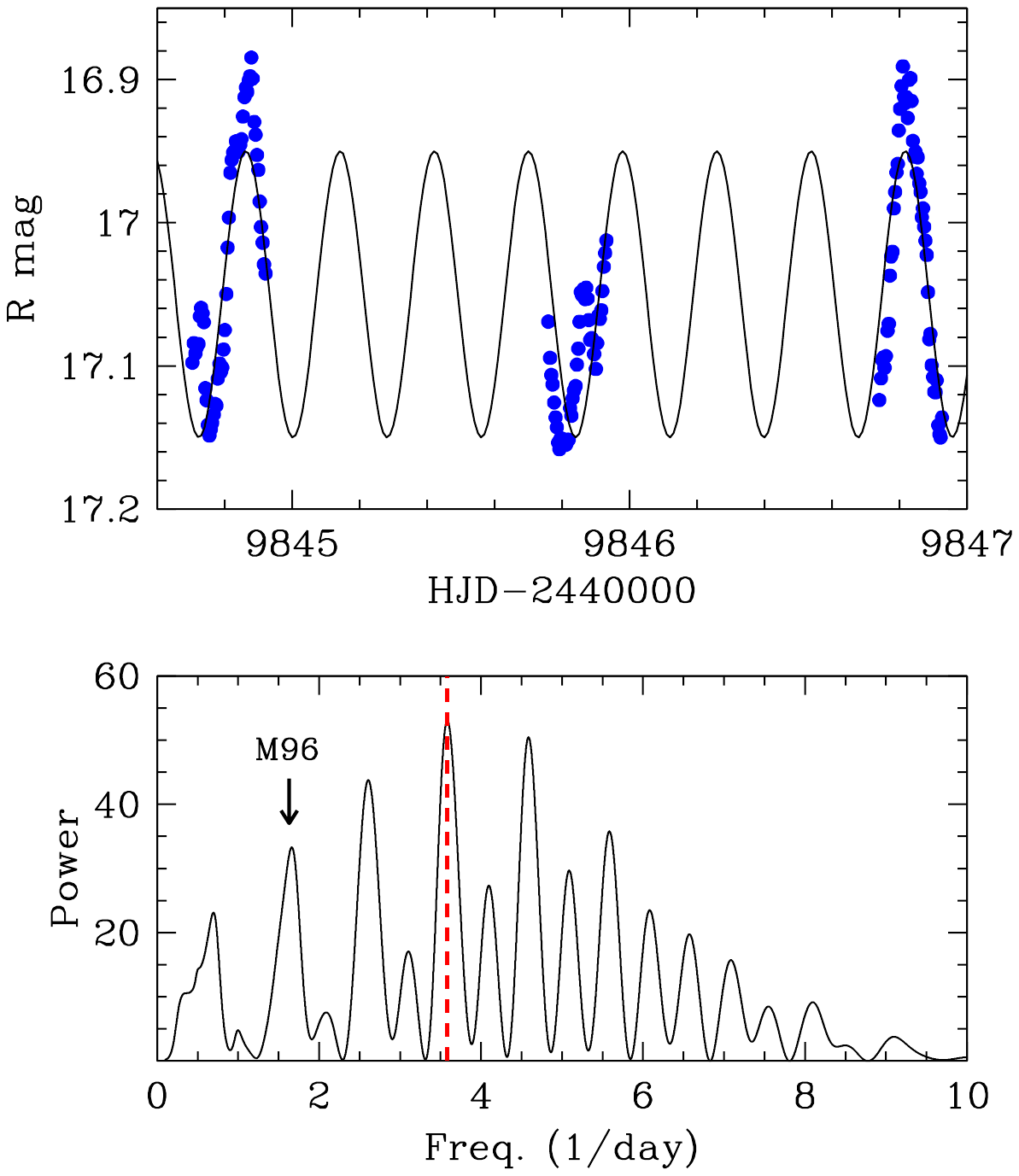}
   \caption{Top panel: Light curve of GRS1716 obtained on the nights of 7-9 May 1995. A sine wave with a 0.279 d 
   (=6.7 h) period is overplotted. Bottom panel: Lomb-Scargle periodogram of the 1995 photometry. The red dashed line indicates 
    the highest  peak, corresponding to 3.584  cycles d$^{-1}$ (=6.7 h). The vertical arrow labels the 14.7 h period 
    reported by M96.       
} 
   \label{fig:fig1}
\end{figure}

\begin{figure}
	\includegraphics[angle=-90,width=\columnwidth]{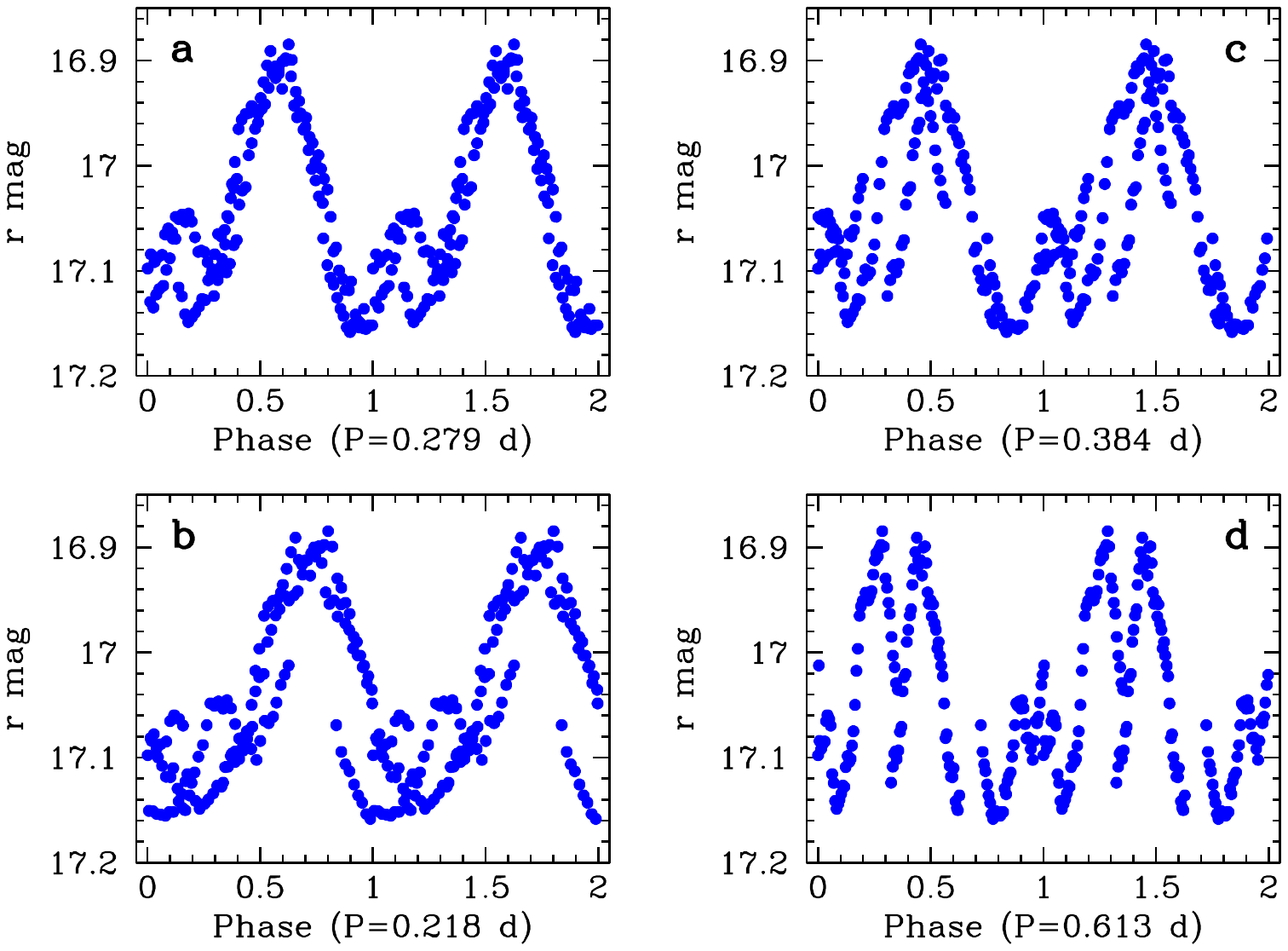}
    \caption{Light curve of GRS1716 folded on the four periods with the highest power in the Lomb-Scargle periodogram. 
    The 0.279 d period (panel a) is clearly favoured by the reduced data dispersion. Panel d displays the data folded on 
    M96's period.       
} 
    \label{fig:fig2}
\end{figure}

Fig.~\ref{fig:fig1} presents the digitized light curve obtained during the 1995 mini-outburst, where a modulation with a timescale of 
a few hours is clearly visible. We have computed a Lomb-Scargle periodogram \citep{lomb76,scargle82} of these data in the 
frequency range 0.1 to 10 cycles d$^{-1}$ at a resolution of 0.001 cycles d$^{-1}$ and the result is plotted in the lower panel 
of  Fig.~\ref{fig:fig1}. The highest peak is found at 3.584 cycles d$^{-1}$ or 6.7 h and it is marked by a vertical dotted 
line. The other main peaks correspond to the 1-day alias pattern introduced by our observing window. 
Note that the 14.7 h  period claimed by M96 (indicated in the figure with an arrow) has significantly 
lower power as it is only the 4th highest peak in the periodogram. 
Inspection of the light curve, folded on the four main periods (Fig.~\ref{fig:fig2} ) strongly favours 0.279 d (=6.7 h) 
as the true period since it is the one that shows minimum dispersion.   
A Gaussian fit to the highest peak in the Lomb-Scargle periodogram yields 0.279$\pm$0.009 d, where we have adopted 
 the sigma of the Gaussian as a conservative estimate of the statistical uncertainty. 

Regarding the discrepancy with M96's period, we note that the latter is based on eight $V$-band night visits 
covering $\sim$20-60 min each (plus two  additional 10-15 min visits), which provides a rather poor phase coverage. The 
14.7 h period is tentatively selected by M96 after substantial data processing through iterative cleaning of the frequencies introduced 
by the window function. Conversely, the 6.7 h period is almost completely sampled on two of our three consecutive visits and 
stands out as the peak with the highest power in the dirty periodogram, without any need for data processing (Fig.~\ref{fig:fig1}). 
Furthermore, the 14.7 h period is clearly disproved by the large scatter shown by our data in the folded light curve. 
All in all we conclude that M96's period is an alias of the 6.7 h modulation and should not be used  
in future studies of GRS1716.
 
The light curve folded on the 0.279 d period exhibits a sawtooth morphology,  
with a small secondary peak around minimum light that moves in phase across nights. Triangular  
modulations, with superimposed (higher frequency) waves are typically observed in outburst light curves of BH XRTs  
and are interpreted as superhumps \citep{odonoghue96,zurita08,thomas22}. 
As a matter of fact, the extreme mass ratios, characteristic of BH XRTs,  tend to prevent donor stars from being exposed 
to X-ray irradiation and, consequently, to exhibit orbitally modulated light curves (see e.g. \citealt{torres21}). 
In any case, since superhump periods are typically $\approx$2 per cent  longer than orbital, the latter (if present) would 
not be resolved from the superhump frequency in our data.
  
 \subsection{Quiescent photometry}
\label{sec:quiescence} 

 \subsubsection{WHT+ACAM 2019 Epoch}
\label{sec:2019} 

\begin{figure*}
\center
\includegraphics[width=7.9cm,angle=0]{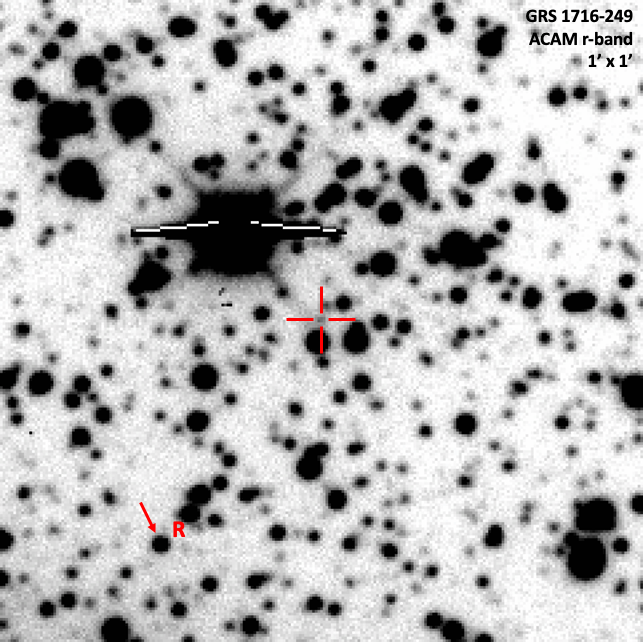}
\includegraphics[width=7.9cm,angle=0]{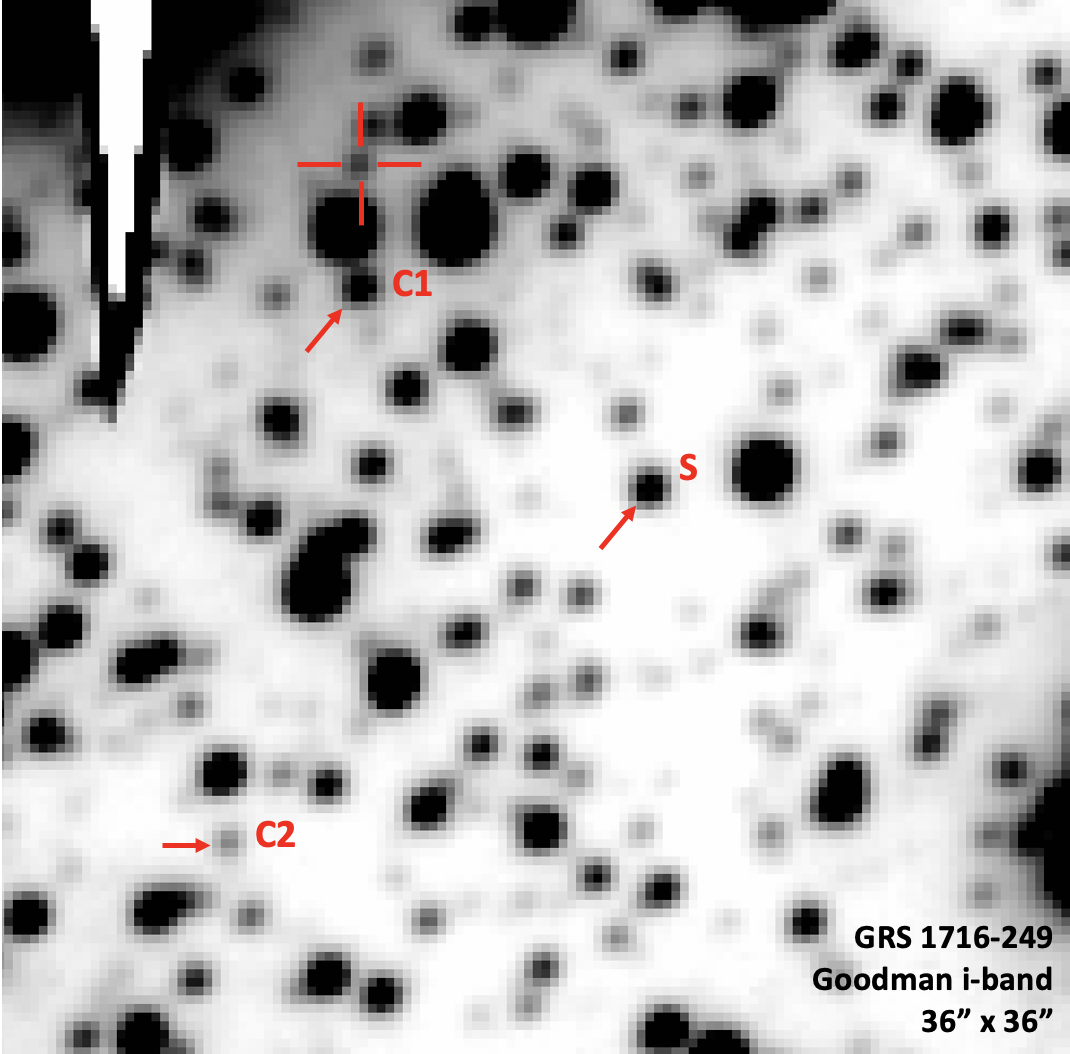}
\caption{Left panel: 600 s r-band image of GRS1716 in quiescence, obtained with ACAM on the night of 2019 July 24. The target is 
indicated by a cross mark, while reference star R with an arrow. North is up and East to the left. 
Right panel:  Zoom in on the average of the 10$\times$600 s i-band images of GRS1716 with seeing $<$1 arcsec, obtained with 
SOAR on the night of 2021 May 31. The target, two comparison stars (C1 and C2) and a local standard (S) are indicated. }
\label{fig:fig3}
\end{figure*}

Fig.~\ref{fig:fig3} (left panel) displays the ACAM Sloan-r band image of the GRS1716 field, obtained on the night of July 24   
2019. The field was calibrated relative to a reference star (labelled R in the figure), selected from the Pan-STARSS 
Data Release 2 catalogue \citep{chambers16}. Its 
magnitude and colour are listed in Table~\ref{tab:tab2}.  
Due to the faintness of the target and the presence of nearby bright stars 
we used point-spread function (PSF) photometry  to measure the brightness of GRS1716 and find $r=22.60\pm0.06$.
A consecutive i-band image, collected only 10 min after, gives $i=21.59\pm0.03$. On the other hand, the mean 
magnitude of  three i-band images obtained on Aug 23 2019 gives $i=22.08\pm0.07$, i.e. a 0.5 mag difference which  
may be ascribed to a combination of flickering, orbital phasing and possible episodic variability between the two visits. 

 \subsubsection{SOAR+Goodman 2021 Epoch}
\label{sec:2021} 

The right panel of Fig.~\ref{fig:fig3} shows a close-up view of the GRS1716 field in the $i$-band, observed with 
SOAR on 2021 May 31. Standard PSF photometry was also performed during the SOAR campaign.  
A Moffat distribution model was constructed for seven selected stars after removing the neighbours. 
This was fitted to the target and a number of close stars in the field,  with their positions fixed to avoid the PSF 
process shifting the position of faint stars towards brighter nearby objects. After several tests, we set the fitting 
and aperture radii to 2 pixels. Differential photometry was performed on the target and two comparison stars 
(labelled as C1 and C2 in Fig.~\ref{fig:fig3}), relative to a local standard (S) whose Pan-STARRS magnitude is 
$i=20.68\pm0.04$. Note that stars brighter than S (e.g. reference star R) were registered outside the detector linearity 
regime. Comparison C2 has similar brightness to the target while C1 was chosen to monitor the possible impact that 
the bright star, located 2.1" SW of the target, could have in the PSF photometry. 
Table~\ref{tab:tab2} summarizes the mean magnitudes of all the reference stars.

 \begin{table}
	\centering
	\caption{Magnitudes of GRS1716 and PS1 reference stars}
	\label{tab:tab2}
	\begin{tabular}{lcc}
		\hline
Star & r & i  \\ 
 		\hline
R    & 20.277$\pm$0.015 & 19.574$\pm$0.015 \\
S    & 21.643$\pm$0.045 & 20.678$\pm$0.037 \\
C1  & 21.710$\pm$0.030 &  20.881$\pm$0.007 \\
C2  &   22.982$\pm$0.003 &  21.968$\pm$0.016 \\
GRS1716$^{\dagger}$ & 23.19$\pm$0.15 & 22.14$\pm$0.14 \\
                 \hline
	\end{tabular}

$^{\dagger}${~We quote the mean magnitudes and rms from the nights of 2021 May 31 and June 1.}

\end{table}

\begin{figure}
	\includegraphics[angle=0,width=\columnwidth]{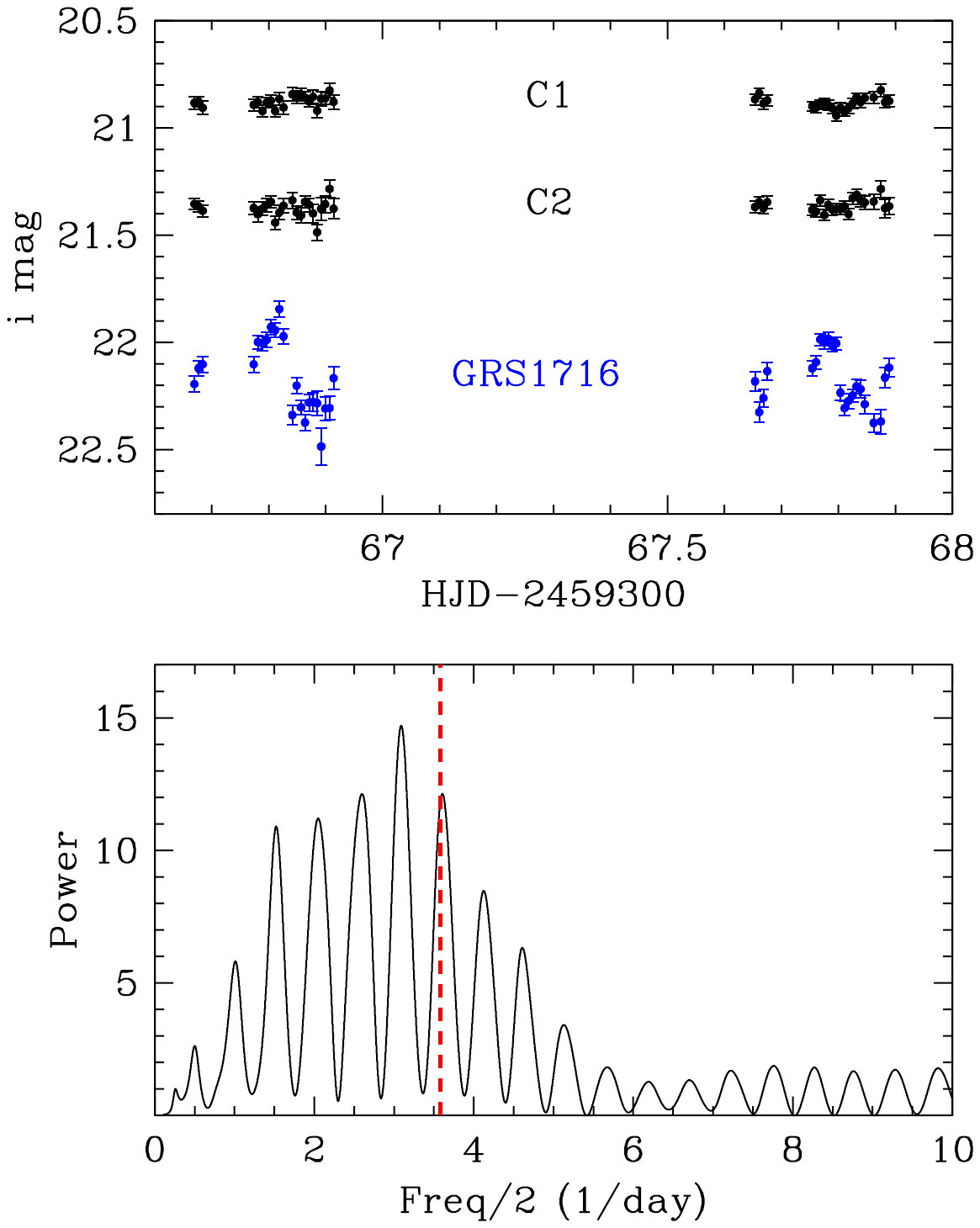}
    \caption{Top panel: SOAR light curve of GRS1716  in quiescence and two comparison stars, obtained on the nights of 2021 
    May 31 and June 1. For the sake of display we have shifted star C2  by --0.6 mag.  
    Bottom panel: Lomb-Scargle periodogram of the GRS1716 light curve. We plot half the frequency in the x-axis 
    because of the double-wave nature of the expected ellipsoidal modulation.The red dashed line indicates 
    the superhump frequency (=6.7 h), which coincides with the second highest peak in the periodogram.       
} 
    \label{fig:fig4}
\end{figure}

Fig.~\ref{fig:fig4} shows the light curve of GRS1716 and stars C1 and C2. Unlike the comparison stars,  GRS1716 
is clearly variable.  As is usually the case in quiescent XRTs, we expect this variability to be dominated by the changing 
visibility of the tidally distorted companion star (the so-called ellipsoidal modulation), with some contamination due to the 
accretion disc light \citep{martin95,shahbaz96}. The ellipsoidal  modulation has two maxima and two minima per orbital cycle 
and its amplitude increases with inclination. 
To determine the orbital period of GRS1716 we have computed a Lomb-Scargle 
period analysis of the SOAR light curve and the result is displayed in the bottom panel of Fig.~\ref{fig:fig4}.  Note that, because of 
the double modulation per orbital cycle  of the ellipsoidal variability we opt for plotting half the frequency in the x-axis. 

\begin{figure}
	\includegraphics[angle=0,width=\columnwidth]{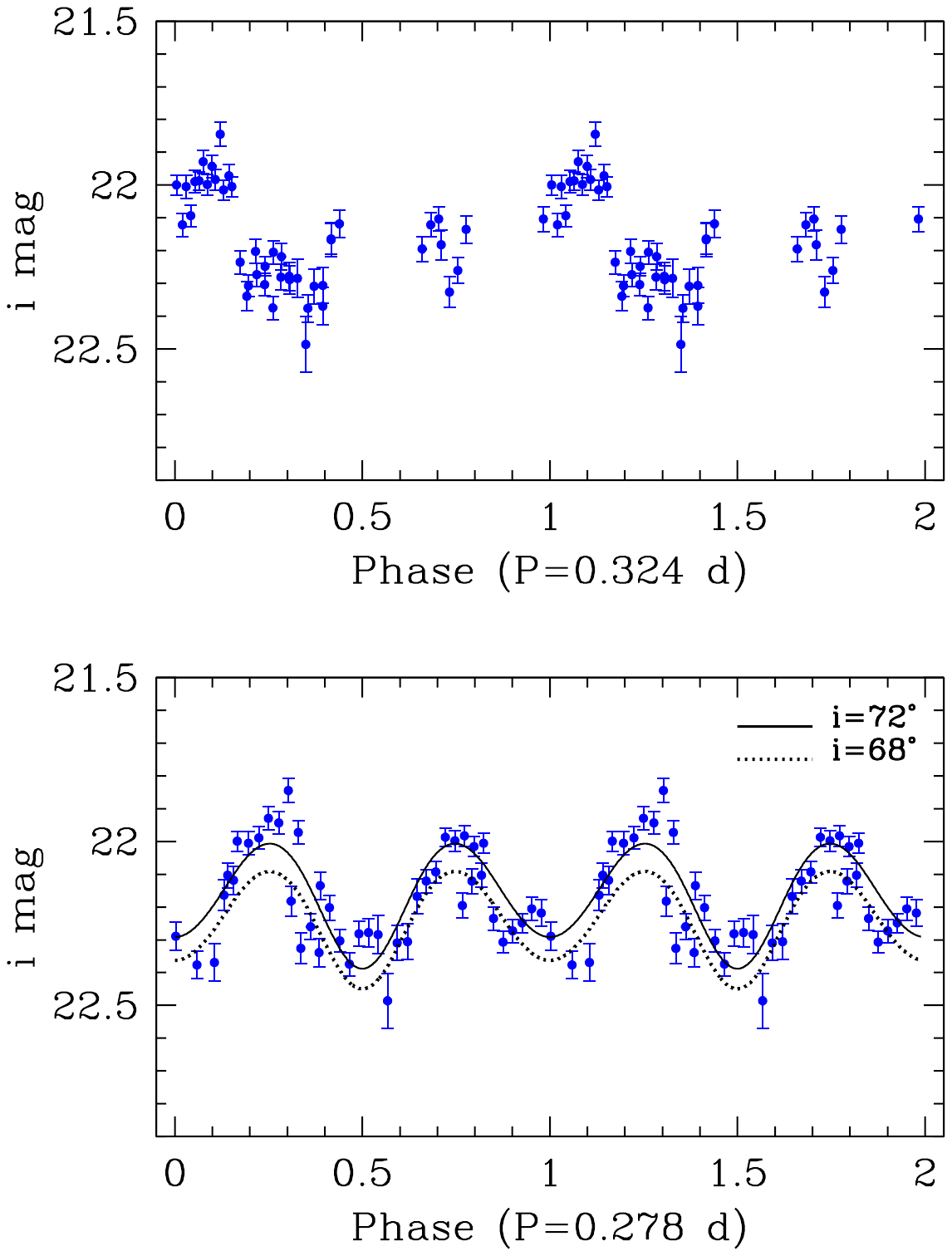}
    \caption{
   Top panel: SOAR light curve of GRS1716 folded on 0.324 d, the period associated with the highest peak in 
   the Fig.~\ref{fig:fig4} periodogram. Bottom panel: same as before, but folded on 0.278 d , our favoured orbital period. 
    The best ellipsoidal model fit with $i=72^{\circ}$ is overplotted in solid line, while an ellipsoidal fit to the 
    lower envelope of the light curve is shown in dashed line style. The latter yields a best inclination $i=68^{\circ}$ 
    (see Sect.~\ref{sec:2021} for details). 
} 
    \label{fig:fig5}
\end{figure}

The highest peak (0.324 d)  is likely an artefact of data sampling because the folded light curve only covers $\sim$60 per 
cent of the orbit (see top panel in  Fig.~\ref{fig:fig5}). In addition, it does not match any of the strong peaks in the 1995 
outburst periodogram (Fig.~\ref{fig:fig1}).   As a matter of fact, the second highest peak in the SOAR periodogram 
coincides with the superhump frequency ($\nu/2$=3.597 cycles day$^{-1}$  or 0.278 d), a strong indication that this is 
the true orbital period ($P_{\rm orb}$). The remaining peaks at shorter frequencies also show significant gaps in 
their phase folded light curves  and we interpret them as sidelobes of the orbital frequency.
Further independent support for the 0.278 d orbital period is provided by the detection of a $\approx$6.5 h modulation in both 
the radial velocities  and the EW of the \ha~line observed during the 2017 outburst (see Appendix~\ref{ap:ha}).  
We hereafter adopt $0.278\pm0.008$ d as the orbital period in GRS1716, where the error corresponds 
again to the sigma of a Gaussian fit to the peak in the periodogram. 

The bottom panel in Fig.~\ref{fig:fig5} displays the GRS1716 quiescent light curve folded on $P_{\rm orb}$=0.278 d.  
We have adopted a tentative zero phase $T_{0}$(HJD)=2459365.963 which places the deepest minimum 
at orbital phase 0.5, referred to as the superior conjunction of the companion star. The reason for a deeper minimum 
lies in the larger gravity darkening of the inner Lagrangian point. 
The large amplitude of the folded light curve and the difference between the two minima suggests that GRS1716 is seen 
at high inclination. In order to constrain the latter we have performed an ellipsoidal model fit
using the code XR$_{\rm BINARY}$ \citep{bayless10}. This is a star-only fit, where we have assumed a 
binary mass ratio $q=0.1$,  $K{_2}=521$  \kms~(Sect.~\ref{sec:spectroscopy}) and 
$T_{\rm eff}=4300$ K (i.e. adequate for a K6V companion, see Sect.~\ref{sec:spectral_type}). 
The flux spectrum of the companion star is computed from the stellar atmosphere models of \cite{kurucz96}, using a 
non-linear limb-darkening law \citep{claret00b} valid for $\log g =0.0-0.5$. The gravity darkening 
depends on the star's effective temperature and is based on \cite{claret00a}. 
The best fit is obtained for $i=72^{+4}_{-5}$ deg, although this 
number should be treated with caution. 
To start with, the ellipsoidal modulation appears strongly contaminated by flickering activity, 
as is commonly observed in the light curves of quiescent XRTs (e.g. \citealt{zurita03}). In addition, its amplitude 
might be diluted  by the (steady) contribution of the disc light, which has so far been neglected.  

In an effort to estimate the impact of flickering in our modeling we have followed \cite{pavlenko96} 
and fitted the lower envelope of the light curve. This is expected to trace the true ellipsoidal variability,  
with minimum flickering contamination. We computed the lower envelope curve through an iterative sigma clipping 
process that rejects data points $>2\sigma$ above a double sine wave model, with periods fixed to $P_{\rm orb}$ 
and $0.5\times P_{\rm orb}$. The latter is a good approximation to the real ellipsoidal modulation, but much less 
expensive in terms of computing time. The process converges after four iterations. A proper ellipsoidal fit to the 
resulting lower envelope curve now yields $i=68\pm 6^{\circ}$ (see bottom panel in Fig.~\ref{fig:fig5}). This exercise 
suggests that flickering variability tends to bias inclinations high while also underestimates the quoted uncertainties.

\subsubsection{Quiescent magnitudes and colour}
\label{sec:colour} 

The mean $i$-band magnitude of the two SOAR nights is $i=22.14\pm0.14$, where the error reflects the standard deviation 
of the light curve. Note that this is significantly fainter than the quiescent magnitude reported in \cite{saikia22}. 
In order to infer the quiescent colour of GRS1716 we examined the  two $r$-band images gathered 
in this run and find $r=23.20\pm0.10$ (2021 May 31) and $r=23.70\pm0.20$ (2021 June 1). Given the large variability 
exhibited by GRS1716 we need to compare these magnitudes with those from the nearest $i$-band images and find   
$(r-i)=1.10\pm0.13$ and $(r-i)=1.30\pm0.22$ respectively. 
These values are consistent within errors with the measurement $(r-i)=1.01\pm0.07$, obtained on 2019 July 24 with ACAM.  
This indicates that, despite intrinsic variability, the colour 
appears to be very stable over the years. 
The weighted mean of the three values, i.e. $(r-i)=1.05\pm0.06$, will be adopted hereafter as 
representative of the quiescent colour of GRS1716. Similarly, we combine this colour with the average $i$-band magnitude 
of the two SOAR nights ($i=22.14\pm0.14$) to derive a mean quiescent $r$-band magnitude of 23.19$\pm$0.15.  

\subsection{Quiescent spectroscopy}
\label{sec:spectroscopy} 

\begin{figure}
	\includegraphics[angle=-90,width=\columnwidth]{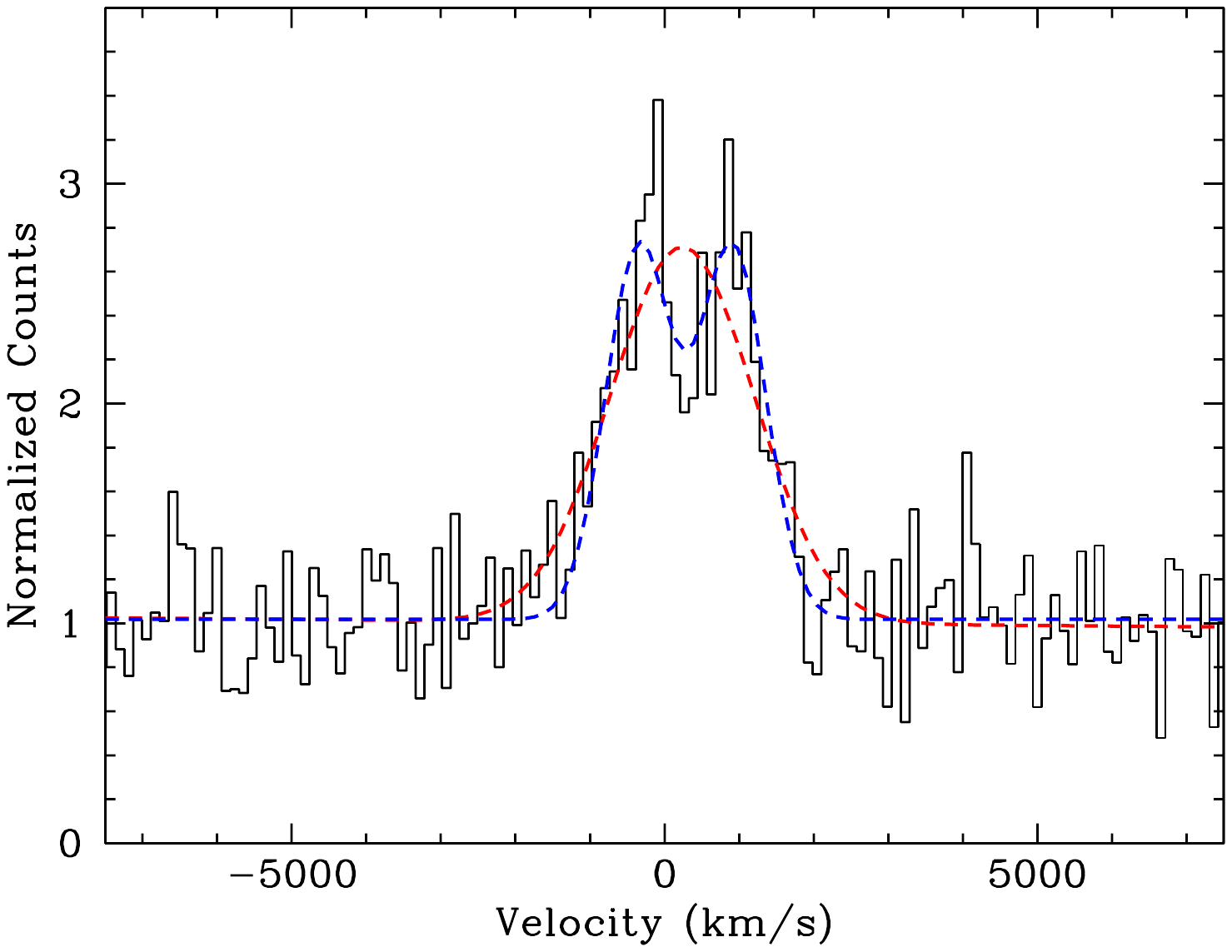}
    \caption{\ha~profile of GRS1716 in quiescence, obtained from the average of four 1800 s GTC spectra. A single 
    Gaussian fit is plotted in red dashed-line style and a symmetric 2-Gaussian model in blue dashed line.        
} 
    \label{fig:fig6}
\end{figure}

The four GTC spectra collected in 2020 have a mean signal-to-noise ratio of $\approx$2 in the continuum. 
The only notable feature in the average spectrum is the presence of a broad double-peaked \ha~emission line, 
characteristic of an accretion disc (Fig.~\ref{fig:fig6}). The line is strong, with a mean equivalent width EW=107$\pm$5 \AA, 
and its centroid velocity is redshifted by $209\pm54$ \kms. Since no absorption lines from the companion are detected 
we need to exploit the properties of the \ha~profile to set constraints on the system parameters. 

First, we use the FWHM-$K_2$ correlation  to determine the radial velocity semi-amplitude of the donor star 
$K_2$ \citep{casares15}. The width of the line is measured by fitting a single Gaussian to the four individual spectra 
in a window of $\pm$10000 \kms~centered on \ha. This yields an average of FWHM=2149$\pm$260 \kms, where the 
quoted errors represent the standard deviation in the distribution of values. Since the individual spectra are rather noisy 
we have also fitted the averaged spectrum and find FWHM=2235$\pm$137 \kms. Observations of quiescent XRTs 
show evidence of accretion disc activity, leading to $\sim$10 per cent variability in FWHM \citep{casares15}.
Given the limited phase coverage ($\sim$25 per cent of the 0.278 d period) we decided to adopt the results of the 
fit to the averaged spectrum, but with a conservative 10 per cent  error to account for possible orbital and episodic 
variability i.e. FWHM=2235$\pm$224 \kms. The empirical scaling $K_2=0.233\times{\rm FWHM}$ thus yields 
$K_{2}=521\pm52$ \kms. 

Second, we use the correlation between $q$ and the ratio of the double peak separation ($DP$) to the line 
width (FWHM) to estimate the former \citep{casares16}. A symmetric 2-Gaussian model fit to the average profile gives 
$DP= 1263\pm49$ \kms, which leads to $q=0.074^{+0.349}_{-0.059}$ (68 per cent confidence through a Monte Carlo 
simulation, see \citealt{casares16}) 
via the equation $\log q = -6.88~ -23.2~\log (DP/{\rm FWHM})$. Unfortunately, the large errors caused by the limited 
quality of the \ha~profile leaves this parameter largely unconstrained. 

Finally, we measure the depth of the \ha~central trough to constrain the inclination as in \cite{casares22}. 
The same 2-Gaussian model yields $W=973\pm72$ \kms~for the full-width-half-maximum of the two fitted 
Gaussians which, together with $DP$,  leads to a normalized double peak trough depth $T=1-2^{1-(DP/W)^2}=0.40\pm0.12$ 
and thus $i=61\pm12^{\circ}$ via equation $i(deg) = 23.7 + 93.5~T$. Here, $T$ has been corrected by 
$\Delta T=0.097\times\Delta V^{2}_{\rm res}/(DP^{2}-W^{2})=+0.02$ in order to account for smearing in the \ha~profile caused by 
our limited spectral resolution $ \Delta V_{\rm res}=378$ \kms. The quoted uncertainty in the inclination has been 
computed through a Monte Carlo simulation as in  \cite{casares22}. 

\section{Discussion}
\label{sec:discussion}

By combining our photometric determination of the orbital period with the $K_2$ velocity of the companion star we 
constrain the mass function of the compact object to 

\begin{equation}
f(M) =  {K_{2}^{3} P_{\rm orb}\over{2 \pi G}} = {M_{1}\sin^{3} i\over{(1+q)^2} }= 4.1 \pm 1.2~M_{\odot}~.
\label{eq:fm}
\end{equation}

\noindent 
This provides strong support for the presence of a BH as it exceeds the $\approx$3 \msun~limit for a stable neutron star.  
The BH nature is consistent with the global X-ray properties, in particular the detection of Type-C QPOs in both the 1993 
and 2017 outbursts \citep{vanderhooft96,bharali19}.  
Note that our mass function relies on a tight FWHM-$K_2$ correlation  that stems from a fundamental scaling between 
the dynamics of the accretion disc and the velocity of the companion star in quiescent XRTs \citep{casares15}. 
Arguably, this is the only avenue to derive dynamical 
information on such a faint ($r\approx$23) target, as traditional methods (i.e. Doppler monitoring of weak metallic lines from 
the companion) are exceedingly time consuming or simply impossible with the available instrumentation. 
For a better knowledge of the BH mass we now need to 
look at other observational constraints that can be enforced on the binary parameters.

\subsection{Restrictions on fundamental parameters}
\label{sec:parameters}

\subsubsection{Spectral type of donor star}
\label{sec:spectral_type}

Under the assumption that the companion star fills its Roche lobe, the orbital period implies a mean stellar density 
$\langle \rho \rangle \approx2.5$ gr cm$^{-3}$ through equation $\langle \rho \rangle \approx110\times 
P_{\rm orb}^{-2}$, where $P_{\rm orb}$ is in units of hours \citep{frank02}. 
Such density is consistent with a $\approx$K5 main sequence star \citep{drilling02}.   
The spectral type of the donor star can, in principle, be constrained through 
the observed quiescent $(r-i)$ colour. Our photometry gives $(r-i)=1.05\pm0.06$ which, after 
correcting for interstellar (IS) extinction in the Sloan bandpasses 
($A_{\rm r}=2.285\times E(B-V)$ and $A_{\rm i}=1.698\times E(B-V)$; see  \citealt{schlafly11}), 
together with our reddening determination $E(B-V)=1.0\pm0.1$  (see Appendix~\ref{ap:extinction}), 
results in $(r-i)_{0}=0.46\pm0.29$. 
This agrees well with a K6 main sequence star \citep{covey07} although, strictly speaking, it only sets 
an upper limit to the spectral type since we have so far neglected any (blue) light contamination from the 
accretion disc. For example, \cite{cantrell10} estimate the fractional contribution of the accretion disc to 
the total light in A0620-00 during the {\it passive-state} (i.e. that with the lowest flux level and variability), both in the 
$V$ and $I$ bands as $f_{\rm V}=0.35$ and $f_{\rm  I}=0.25$, respectively. 
A linear interpolation to the effective wavelength of the Sloan 
filters yields $f_{\rm r}=0.32$ and $f_{\rm i}=0.27$, which leads to excess magnitudes $\Delta r=0.42$ and 
$\Delta i=0.34$. If our GRS1716 magnitudes were affected by a similar amount of disc contamination as in the 
passive-state of A0620-00, the {\it dereddened} $(r-i)$ colour would rise by $\sim$0.08 mag, thus favouring a $\sim$K7 
spectral type. It should be borne in mind that quiescent BH XRTs are sometimes caught in the so-called 
{\it active-state}, with higher levels of disc activity, although we note that this does not seem to have a large impact 
on the observed colour \citep{cantrell08,dincer18}. 
In conclusion, after correcting for extinction and disc veiling, we find that the quiescent $(r-i)$ colour of GRS1716 
favours a $\approx$K6-7 spectral type, even though we caution that large uncertainty is associated with the knowledge of IS 
reddening. Incidentally, this classification is in excellent agreement with the spectral type predicted by the empirical 
spectral type-period relationship of \cite{smith98} for $P_{\rm orb}$=6.7 h. 

\subsubsection{Mass ratio}
\label{sec:mass_ratio}

The evidence for superhumps in 1995 implies that the accretion disc expanded beyond 
the tidal 3:1 resonance radius and, therefore, $q\lesssim$0.25 \citep{paczynski77}. 
An independent constraint is placed by the spectral classification of the companion star.  
Evolutionary models  of  short period ($\lesssim$12 h) X-ray binaries provide an 
upper limit to the mass of the donor star based on zero-age main sequence stars 
with the same $T_{\rm eff}$ \citep{kolb01}. 
Our K6 V classification thus implies M$_{2}\lesssim0.69$ \msun~and, since 
the compact object is a BH with M$_{1}\gtrsim3$ \msun, then $q\lesssim0.23$.  
A tentative mass ratio can in fact be derived from the radial velocities of the \ha~line during outburst. 
We show in Appendix~\ref{ap:ha} that the red peak of the line on 2017 June 3 traces a 
sine wave with a semi-amplitude of 79 \kms. Under the assumption that this reflects the 
motion of the compact star we derive $q\approx79/521=0.15$,  
although we warn about possible biases in the radial velocity curve caused by contamination from a putative 
superhump (e.g. \citealt{zurita02,torres02,haswell04}). 
Finally, our 2-Gaussian fit to the quiescent GTC spectrum gives 
a very loose constraint $q=0.015-0.423$ (68 per cent confidence). 
It should be noted that large $q$ values, such as 0.423, can in fact be ruled out because, when combined with 
 M$_{1}\gtrsim3$ \msun, imply  M$_{2}\gtrsim1.3$ \msun~i.e. a F5 V star that would not fit in the Roche lobe of a 
 6.7 h orbit. Furthermore, the ensemble of $q$ values for BH XRTs with low-mass companions 
are narrowly distributed around $q=0.06$, and none exceeds $q=0.18$ \citep{casares16, heida17, torres20}. 
To summarize on the mass ratio restrictions, given the information that is available from the photometric evidence 
of superhumps, the orbital period, the \ha~velocities and double peak fit we 
decide to adopt the wide range $q=0.02-0.25$. 

\subsubsection{Inclination}
\label{inclination}

The 2017 outburst of GRS1716 was intensively scrutinized by {\it Swift} and {\it NuSTAR}, with no X-ray eclipses nor 
dips being  detected \citep{bharali19}. This sets a rough upper limit to the binary inclination 
$i\lesssim70^{\circ}$ \citep{frank87}. On the other hand,  the observation of double-peaked \ha~profiles in quiescence 
and outburst suggests $i\gtrsim50^{\circ}$ (see Fig~\ref{fig:fig6} and Appendix~\ref{ap:ha}). 
Further, by fitting the depth of the double peak trough in the quiescent GTC profile we measure $i=61\pm12^{\circ}$ (Sect.~\ref{sec:spectroscopy}). This agrees well with the inclination obtained for XTE J1859+226 
($i=66.3\pm4.3^{\circ}$), a {\it twin} of GRS1716  with identical $P_{\rm orb}$ and \ha~line properties \citep{corral11,yanes-rizo22}. 
 In particular, both lines are equally strong (EW=102$\pm$12 \AA~and $107\pm11$ \AA, 
 respectively), a fact that suggests small foreshortening of the disc continuum light and, therefore, a 
 moderately high inclination in the two systems \citep{warner86}. 
A relatively high inclination $i=72^{+4}_{-5}$ deg is favoured by the ellipsoidal fit to the quiescent $i$-band light curve, 
although systematics caused by flickering might be important. 
Circumstantial evidence for a relatively high inclination is also provided by the large disc wind velocities  
observed by \cite{cuneo20} during the 2016-2017 outburst (see Table 2 in \citealt{panizo22}).
Given all the above we decide to take the conservative value 
$i=61\pm15^{\circ}$, where we have adopted a larger uncertainty than provided by the fit to the 
\ha~trough in order to encompass the inclination range allowed by the ellipsoidal modeling (Sect.~\ref{sec:2021}).

\subsection{Black hole mass}
\label{sec:mass}

In order to estimate the BH mass in GRS1716 we have run a Monte Carlo simulation with 10$^5$ trials on eq.~\ref{eq:fm} 
and the constraints above. In summary, we take Gaussian distributions for $K_2$ and $P_{\rm orb}$ 
with mean and standard deviations set by our observed values i.e. $K_{2}=521\pm52$ \kms~and 
 $P_{\rm orb}=0.278\pm$0.008 d. We also adopt a normal distribution of inclination angles with $i=61\pm15^{\circ}$, but 
 truncated at $i=90^{\circ}$. Regarding the mass ratio we assume a uniform distribution between $q=0.02-0.25$. 
 This yields $M_{1}=8.3^{+7.4}_{-3.1}$ \msun~and $M_{2}=1.1^{+1.4}_{-0.7}$ \msun~for the masses of the two binary components. 
We note, however, that most $M_2$ solutions are ruled out by our spectral type determination and, therefore, decided 
to enforce further ad hoc restrictions. 

First, from the lower limit of our quiescent de-reddened colour $(r-i)_{0}=0.46\pm0.29$ 
we set a conservative upper limit to the donor's spectral type of G4V and thus $M_{2}<0.98$ \msun. 
Subsequently, given the absence of X-ray eclipses we impose a more restrictive upper limit on inclination based on the geometrical 
limit $\cos i < R_{2}/a$, where the ratio between the donor star radius $R_2$ and binary separation $a$ is tied to $q$ by 
Eggleton's expression \citep{eggleton83}. With these additional constraints we find  $M_{1}=6.4^{+3.2}_{-2.0}$ \msun~and 
$M_{2}=0.5\pm0.3$ \msun~through the Monte Carlo analysis. 

The large fractional uncertainty in $K_2$ and, especially, the loose 
 constraint on inclination lead to a BH mass distribution that is very skewed towards high values. 
 Tighter constraints on these two parameters are thus fundamental to obtain a more accurate BH mass. 
An independent  determination of the BH mass can also be obtained by simply combining the intrinsic width of the double peak 
profile ($W$) with $P_{\rm orb}$ through  equation 8 in \cite{casares22} 

\begin{equation}
M_{1}=3.45\times10^{-8}~P_{\rm orb}~[\left(0.63~W+145\right)/\beta]^{3}~~~~~M_{\odot}, 
\label{eq:mass}
\end{equation}

\noindent 
where $\beta=0.84$ accounts for the fact that the outer disc material is $\approx$84 per cent sub-Keplerian.
This yields $M_{1}=7.0\pm1.5$ \msun, where, following 
\cite{casares22},  we adopt a 20 per cent error in $M_{1}$, based on the distribution of differences with respect to 
the dynamical masses of a  sample of calibrators. 

\subsection{Distance, Galactic elevation and X-ray luminosity}
\label{sec:distance}

Our quiescent $r$-band magnitude and orbital period allow us to constrain the luminosity of GRS1716 and its distance. 
In order to do so we apply the empirical $M_{\rm r}-P_{\rm orb}$ correlation obtained for quiescent BH XRTs in 
\cite{casares18} (eq. 8):

\begin{equation}
M_{r}= 4.64(0.10)-3.69(0.16) \log P_{\rm orb}~(d), 
\label{eq:magnitude}
\end{equation}

 \noindent
For $P_{\rm orb}=0.278$ d we get $M_{\rm r}=6.7\pm0.2$ which, in turn, is fully consistent with 
the absolute $r$-band magnitude of a K5-7 main sequence star. 
The distance to GRS1716 is obtained by comparing $M_{\rm r}$ with the observed magnitude $r=23.19\pm0.15$ 
through the usual distance modulus equation 

 \begin{equation}
\left(\frac{d}{\rm kpc}\right)=10^{~\left[0.2~\left(r-M_{r}-A_{r} ~\right)-2\right]}
   \label{eq:distance}
 \end{equation}

\noindent
We here adopt an $r$-band extinction $A_{\rm r}=2.285~E(B-V)$ \citep{schlafly11}, which  
assumes a traditional reddening law $R_{\rm V}=A_{\rm V}/E(B-V)=3.1$, appropriate 
for the diffuse interstellar medium \citep{cardelli89,fitzpatrick99}. 
From the strength of IS lines in the 2017 outburst spectra we find $E(B-V)=1.0\pm0.1$ (see Appendix~\ref{ap:extinction}) 
which results in  $A_{\rm r}=2.29\pm0.23$ and 
a distance $d=6.9\pm1.1$ kpc. We note that this is a factor $\approx3$ 
larger than the 2.4$\pm$0.4 kpc reported in \cite{dellavalle94}, 
but agrees well with the distance favoured by \cite{saikia22}. 
For an independent test on our distance we have looked at 3D reddening maps constructed from 
Pan-STARRS/2MASS photometry and GAIA parallaxes \citep{green19}. These show a steady rise in extinction up to 
2.5--3.0 kpc, where it  jumps to $A_{\rm r}=2.617~E(g-r)=2.4$. Beyond this point, the extinction level remains 
constant all the way up to $\approx$6.5 kpc. 
We therefore conclude that our estimated distance and reddening values are broadly consistent with the available 3D 
reddening maps although unfortunately the latter are not very informative beyond $\gtrsim$3 kpc. 

Our distance determination has wide-ranging implications. 
To start with, it places the binary at $z\sim$0.8 kpc above the Galactic Plane, in better agreement with an observed   
$P_{\rm orb}-z$ anticorrelation which lends support to the existence of natal kicks 
\citep{gandhi20}. In relation to this, \cite{atri19} estimated a {\it potential} kick velocity (i.e. the peculiar velocity when 
the system crosses the Galactic Plane) for GRS1716 of 67$\pm$21 \kms. This was based on its VLBI proper motion, 
vertical elevation and the Galactic rotation velocity for an assumed distance of 2.4 kpc. 
Our revised distance and z-elevation, 
together with a tentative systemic velocity of $209\pm54$ \kms~(derived from the centroid of the quiescent \ha~line), 
implies a larger Galactic Plane crossing velocity in the range $\sim$153-355 \kms~(90 per cent confidence; P. Atri, 
private communication), lending further support for a strong natal kick in GRS1716. 
Next, the new distance raises the peak X-ray and radio luminosities by a factor $\sim$8.3. 
Although these still keep GRS1716 on the radio quiet branch of the L$_{\rm R}$-L$_{X}$ diagram \citep{bassi19, saikia22},  
the new peak X-ray luminosity 
during the 2016-2017 outburst becomes  L$_{\rm X}$(0.5-10 keV)=4.4$\times$10$^{37}$ ergs s$^{-1}$.This 
corresponds to $\approx6$\% L$_{\rm Edd}$ (for a 6 \msun~BH), in good agreement with the linear 
$P_{\rm orb}-L_{\rm X}$ correlation proposed by \cite{wu10}. Such luminosity is also more consistent with the presence 
of a thermal disc component, detected during the hard-intermediate state, and the evidence that GRS1716 attempted 
a failed transition to the soft-state \citep{armas17,bassi19,jiang20}. 
In fact, hard-to-soft state transitions in BH XRTs are typically observed at $\approx$0.1 L$_{\rm Edd}$ 
but rarely at $\sim10^{-3}$ L$_{\rm Edd}$ \citep{maccarone03,yu09,dunn10}, as would be implied if the distance 
were 2.4 kpc. 
A larger distance also shifts GRS1716 from the neutron star to the BH region in the optical/X-ray correlation plots 
(see \citealt{saikia22}).  And finally, the new distance 
will surely affect the accretion disc parameters derived through X-ray 
spectral modeling, such as the mass accretion rate, disc inclination and inner disc radius.  
For example, it increases the mass accretion rate $\dot{m}$ (=L$_{\rm 0.1-100 kev}$/0.2~L$_{\rm Edd}$) 
by one order of magnitude, thus helping to reconcile the low disc density measured by \cite{jiang20} with that of 
GX 339-4 (at similar $\dot{m}$). Our restrictions will certainly 
have an impact too on the BH spin inferred from spectral fitting \citep{tao19}.    
Clearly new analysis of the 2017 X-ray spectra should be performed  
in the light of the new BH mass and distance derived in this paper. 

\section{Conclusions}
\label{sec:conclusions}

We have presented evidence for a 0.278$\pm$0.008 d orbital period in the BH XRT GRS 1716-249, 
based on 1995 superhump variability and ellipsoidal light curves obtained in quiescence. 
The extreme faintness of the optical counterpart ($r=23.19\pm0.15$) prevents a direct detection of spectral 
features from the companion star in quiescence. However, GTC spectroscopy reveals a strong double-peaked 
\ha~emission line that we exploit to derive fundamental binary parameters through a set of scaling relations 
\citep{casares15, casares16,casares22}.  Using these, combined with light curve information, leads to  
$f(M)=4.1\pm1.2$ \msun, $q=0.02-0.25$ and 
$i=61\pm15^{\circ}$. By including further restrictions on the companion's spectral type and the lack of X-ray eclipses, 
we constrain the BH mass to $M_{1}=6.4^{+3.2}_{-2.0}$ \msun. 
Finally, we find that GRS 1716-249 is located 6.9$\pm$1.1 kpc away and 0.8 kpc above the Galactic Plane, 
in support for a strong natal kick. Our new distance implies that the 2017 failed transition to the soft state 
occurred at $\approx6$\% L$_{\rm Edd}$, in agreement with observations of other BH XRTs. 

\section*{Acknowledgements}

We thank the anonymous referee for a useful and constructive report.  
We also kindly thank Pikky Atri for computing the potential kick velocity of GRS 1716-249. 
We are grateful to Frank van der Hooft for reducing the 1995 photometry and producing the light curves used in this paper. 
This article is based on observations made at the Observatorio del Roque de los Muchachos with the William Herschel 
Telescope (WHT) and the Gran Telescopio Canarias (GTC) operated on the island of La Palma by the IAC. The GTC observations 
were obtained under Director's Discretionary Time GTC/WHT/2020-153 of Spain's Instituto de Astrof\'isica de Canarias. 
Also based on observations obtained at the Southern Astrophysical Research (SOAR) telescope, which is a joint project of the 
Minist\'{e}rio da Ci\^{e}ncia, Tecnologia e Inova\c{c}\~{o}es (MCTI/LNA) do Brasil, the US National Science Foundation’s NOIRLab, 
the University of North Carolina at Chapel Hill (UNC), and Michigan State University (MSU). The digitized light curves were based 
on observations made at the European Southern Observatory, La Silla, Chile. This work is supported by the Spanish
Ministry of Science under grants PID2020-120323GB-I00, PID2021-124879NB-I00 and EUR2021-122010.
MAP also acknowledges  support from the Consejer\'\i{}a de Econom\'\i{}a, Conocimiento y Empleo 
del Gobierno de Canarias and the European Regional Development Fund (ERDF) under grant with reference 
ProID2021010132. KM is funded by the EU H2020 ERC grant no. 758638. 
Molly software developed  by  Tom  Marsh  is  gratefully  acknowledged. 
We dedicate this paper to the memory of Jan van Paradijs, whose contributions to the field of X-ray binaries have 
been a source of inspiration. Jan was also the PI of the 1.5m Danish 1995 proposal that provided 
the evidence for a 6.7 h superhump.  

\section*{Data availability}
The data used in this article (except for the Danish 1.54-m light curves) is publicly available from the 
relevant observatory archives. All data will be shared on reasonable request to the corresponding author.





\begin{thebibliography}{99}


\bibitem[Armas Padilla \& Mu\~noz-Darias (2017)]{armas17}
Armas Padilla M, Mu\~noz-Darias T., 2017, The Astronomer's Telegram, 10236

\bibitem[Atri et al.(2019)]{atri19}
Atri P. et al., 2019, \mnras, 489, 3116

\bibitem[Ballet et al.(1993)]{ballet93}
Ballet J., Denis M., Gilfanov M., Sunyaev R.A., 1993, IAU Circ. 5784

\bibitem[Bassi et al.(2019)]{bassi19}
Bassi T. et al., 2019, \mnras, 482, 1587

\bibitem[Bassi et al.(2020)]{bassi20}
Bassi T. et al., 2020, \mnras, 494, 571

\bibitem[Bayless et al.(2010)]{bayless10}
Bayless  A.J., Robinson E.L., Hynes R.I., Ashcraft T.A., Cornell M.E.,, 2010, \apj, 709, 251

\bibitem[Belloni et al.(2011)]{belloni11}
Belloni T.M., Motta S.E., Mu\~noz-Darias T., 2011, {\it Bull. Astr. Soc. India}, 39, 409

\bibitem[Bharali et al.(2019)]{bharali19}
Bharali P.,  Chandra S., Chauhan J. Garc\'\i{}a J.A., Roy J., Boettcher M., Boruah K., 2019, \mnras, 487, 3150

\bibitem[Borozdin et al. (1994)]{borozdin94}
Borozdin K., Alexandrovich N., Arefiev V., Sunyaev R., 1994, IAU Circ. 6083

\bibitem[Borozdin et al. (1995)]{borozdin95}
Borozdin K., Alexandrovich N., Sunyaev R., 1995, IAU Circ. 6141

\bibitem[Cantrell et al.(2008)]{cantrell08}
Cantrell A.G., Bailyn C.D., McClintock J.E., Orosz J.A., 2008, \apj, 673, L159 

\bibitem[Cantrell et al.(2010)]{cantrell10}
Cantrell A.G. et al., 2010, \apj, 710, 1127 

\bibitem[Cardelli, Clayton \& Mathis(1989)]{cardelli89}
Cardelli J.A., Clayton G.C., Mathis J.S., 1989, \apj, 345, 245 

\bibitem[Casares \& Jonker(2014)]{casares-jonker14}
Casares J., Jonker P.G., 2014, \ssr, 183, 223

\bibitem[Casares (2015)]{casares15}
Casares J., 2015, \apj, 808, 80 

\bibitem[Casares (2016)]{casares16} 
Casares J., 2016, \apj, 822, 99 

\bibitem[Casares (2018)]{casares18}
Casares J., 2018, \mnras, 473, 5195 

\bibitem[Casares et al. (2022)]{casares22}
Casares J. et al., 2022, \mnras, 516, 2023 

\bibitem[Cepa et al. (2000)]{cepa00}
Cepa J. et al., 2000, in Iye M., Moorwood A. F., eds, Proc. SPIE Conf. Ser.
Vol. 4008, Optical and IR Telescope Instrumentation and Detectors. SPIE,
Bellingham, p. 623

\bibitem[Chambers et al. (2016)]{chambers16}
Chambers  K.C. et al., 2016, {\it arXiv:161205560} 

\bibitem[Claret (2000a)]{claret00a}
Claret  A., 2000a, A\&A,  359, 289 

\bibitem[Claret (2000b)]{claret00b}
Claret  A., 2000b, A\&A,  363, 1081 

\bibitem[Churazov et al. (1994)]{churazov94}
Churazov E., Gilfanov M., Ballet J., Jourdain E., 1994, IAU Circ. 6083

\bibitem[Corral-Santana et al.(2011)]{corral11}
Corral-Santana J.M., Casares J., Shahbaz T., Zurita C., Mart\'\i{}nez-Pais I.G., Rodr\'\i{}guez-Gil P., 2011, \mnras, 413, L15 

\bibitem[Corral-Santana et al.(2016)]{corral16}
Corral-Santana J.M., Casares J., Mu\~noz-Darias T., Bauer F.E., Mart\'\i{}nez-Pais I.G., Russell D.M., 2016, A\&A, 587, A61

\bibitem[Covey et al. (2007)]{covey07}
Covey K.R. et al., 2007, AJ, 134, 2398

\bibitem[C\'uneo et al. (2020)]{cuneo20}
C\'uneo V.A. et al., 2020, \mnras, 498, 25

\bibitem[Della Valle et al.(1994)]{dellavalle94}
Della Valle M., Mirabel I.F., Rodriguez L.F.,1994, A\&A, 290, 803

\bibitem[Din\c cer et al.(2018)]{dincer18}
Din\c cer T., Bailyn C.D., Miller-Jones J.C.A., Buxton M., MacDonald R.K.D., 2018, \apj, 852, 4

\bibitem[Drilling \& Landolt(2002)]{drilling02}
Drilling J.S., Landolt A.U., 2002, {\it Normal Stars} in Allen's Astrohysical Quantities, Cox A.N. ed., Springer New York , NY,  p. 381, 
ISBN 978-0-387-95189-8

\bibitem[Dunn et al.(2010)]{dunn10}
Dunn R.J.H., Fender R.P., K\"{o}rding E.G., Belloni T., Cabanac C., 2010, \mnras, 403, 61

\bibitem[Eggleton(1983)]{eggleton83}
Eggleton P.P., 1983, \apj, 268, 368
 
\bibitem[Fitzpatrick (1999)]{fitzpatrick99}
Fitzpatrick E.L., 1999, \pasp, 111, 63

\bibitem[Frank et al.(1987)]{frank87}
Frank J., King A.R., Lasota, J.-P., 1987, A\&A,178, 137

\bibitem[Frank et al.(2002)]{frank02}
Frank J., King A.R., Raine D.J., 2002, Accretion Power in Astrophysics, Vol.
21 (3rd ed.; Cambridge: Cambridge Univ. Press)

\bibitem[Gandhi et al.(2020)]{gandhi20}
Gandhi P., Rao A., Charles  P.A., Belczynski K., Maccarone T.J., Arur K., Corral-Santana J.M., 2020, \mnras, 496, L2

\bibitem[Green et al.(2019)]{green19}
Green G.M., Schlafly E., Zucker C., Speagle J.S., Finkbeiner D., 2019, \apj, 887, 93

\bibitem[G\"{u}ver  \& \"{O}zel (2009)]{guver09}
G\"{u}ver T., \"{O}zel F., 2009, \mnras, 400, 2050

\bibitem[Harmon et al.(1993)]{harmon93}
Harmon B.A., Fishman G.J., Paciesas W.S., Zhang S.N., 1993, IAU Circ. 5874 

\bibitem[Harmon et al.(1994)]{harmon94}
Harmon B.A., Zhang S.N., Paciesas W.S., Wilson C.A., Fishman G.J., 1994,  IAU Circ. 5874 

\bibitem[Haswell et al.(2004)]{haswell04}
Haswell C.A. et al., 2004, Rev Mex Astron Astrofis, 20, 160

\bibitem[Heida et al. (2017)]{heida17}
Heida M., Jonker P.G., Torres M.A.P., Chiavassa A., 2017, \apj, 846, 132

\bibitem[Herbig (1975)]{herbig75}
Herbig H.G., 1975, \apj, 196, 129

\bibitem[Hjellming et al.(1996)]{hjellming96}
Hjellming R.M., Rupen M.P., Shrader C.R., Campbell-Wilson D., Hunstead R.W., McKay D.J., 1996, \apj, 470, L105

\bibitem[Jiang et al.(2020)]{jiang20} 
Jiang J., F\"{u}rst F., Walton D.J., Parker M.L., Fabian A.C., 2020, \mnras, 492, 1947

\bibitem[Kiel \& Hurley (2006)]{kiel06}
Kiel P.D., Hurley J.R., 2006, \mnras, 369, 1152
 
\bibitem[Kolb et al. (2001)]{kolb01}
Kolb U., King A.R., Baraffe I., 2001, \mnras, 321, 544
  
\bibitem[Kurucz (1996)]{kurucz96}
Kurucz R.L., 1996, in Adelman S. J., Kupka F., Weiss W. W., eds, 
Astronomical Society of the Pacific Conference Series Vol. 108, M.A.S.S., 
Model Atmospheres and Spectrum Synthesis. p. 2

\bibitem[Lomb(1976)]{lomb76}
Lomb N.R., 1976, \apss, 39, 447

\bibitem[Maccarone (2003)]{maccarone03}
Maccarone T.J., 2003, A\&A, 409, 697

\bibitem[Marsh(1989)]{marsh89}
Marsh T.R., 1989, \pasp, 101, 1032

\bibitem[Martin et al. (1995)]{martin95}
Martin A.C., Casares J., Charles P.A., van der Hooft F., van Paradijs J., 1995, \mnras, 274, L46

\bibitem[Masetti et al.(1996)]{masetti96} 
Masetti N., Bianchini A., Bonibaker J., Della Valle M., Vio R.,1996, A\&A, 314, 123

\bibitem[Masumitsu et al.(2016)]{masumitsu16} 
Masumitsu, T. et al., 2016, The  Astronomer’s Telegram, 9895 

\bibitem[Mata S\'anchez et al.(2015)]{mata15}
Mata S\'anchez D, Mu\~noz-Darias T., Casares J., Corral-Santana J.M., Shahbaz T., 2015, \mnras, 454, 2199

\bibitem[McClintock \& Remillard (2006)]{mcclintock06}
McClintock J.E., Remillard R.A., 2006, in {\it Compact stellar X-ray sources},  ed. W. Lewin \& M. van der Klis, 
Cambridge Astrophysics Series No. 39, Cambridge University Press, p.157
     
\bibitem[Munari \& Zwitter(1997)]{munari97}
Munari U., Zwitter T. 1997, A\&A, 318, 269

\bibitem[Negoro et al.(2016)]{negoro16} 
Negoro, H. et al., 2016, The  Astronomer’sTelegram, 9876

\bibitem[O'Donoghue \& Charles (1996)]{odonoghue96} 
O'Donoghue D., Charles P.A., 1996, \mnras, 282, 191

\bibitem[Paczy\'nski(1977)]{paczynski77} 
Paczy\'nski B., 1977, \apj, 216, 822

\bibitem[Panizo-Espinar et al.(2022)]{panizo22} 
Panizo-Espinar G. et al., 2022, A\&A, 664, 100

\bibitem[Pavlenko et al. (1996)]{pavlenko96}
Pavlenko E.P., Martin A.C., Casares J., Charles P.A., Ketsaris N.A., 1996, \mnras, 281, 1094

\bibitem[Revnivtsev et al. (1998)]{revnivtsev98} 
Revnivtsev M. et al. , 1998, A\&A, 331, 557

\bibitem[Saikia et al.(2022)]{saikia22}
Saikia  P. et al.,  2022, \apj, 932, 38

\bibitem[Scargle (1982)]{scargle82}
Scargle  J.D, 1982, \apj, 263, 835

\bibitem[Schlafly \& Finkbeiner(2011)]{schlafly11}
Schlafly E.F., Finkbeiner D.P., 2011, \apj, 737, 103

\bibitem[Shahbaz et al.(1996)]{shahbaz96}
Shahbaz T., van der Hooft F., Charles P.A., Casares J., van Paradijs J., 1996, \mnras, 282, L47

\bibitem[Shaw et al.(2016)]{shaw16}
Shaw A.W., Charles P.A., Casares J., Hern\'andez Santisteban J.V., 2016, \mnras, 463, 1314

\bibitem[Smith \& Dhillon(1998)]{smith98}
Smith D.A., Dhillon V.S., 1998, \mnras, 301, 767

\bibitem[Stetson(1987)]{stetson87}
Stetson P.B., 1987, \pasp, 99, 191

\bibitem[Tao et al.(2019)]{tao19}
Tao L., Tomsick J.A., Qu J., Zhang S., Zhang S,, Bu Q., 2019, \apj, 887, 184

\bibitem[Thomas et al.(2022)]{thomas22}
Thomas J.K., Charles P.A., Buckley D.A.H., Kotze M.M., Lasota J.-P., 
Potter S.B., Steiner J.F., Paice J.A., 2022, \mnras, 509, 1062

\bibitem[Tonry et al.(2012)]{tonry12}
Tonry J.L. et al., 2012, \apj, 750, 99

\bibitem[Torres et al.(2002)]{torres02}
Torres M.A.P. et al., 2002, \apj, 569, 423

\bibitem[Torres et al.(2020)]{torres20}
Torres M.A.P., Casares J., Jim\'enez-Ibarra F., \'Alvarez-Hern\'andez A.,  Mu\~noz-Darias T., 
Armas Padilla M., Jonker P.G., Heida M., 2020, \apj, 893, L37

\bibitem[Torres et al.(2021)]{torres21}
Torres M.A.P., Jonker P.G., Casares J., Miller-Jones J.C.A., Steeghs D., 2021, \mnras, 501, 2174

\bibitem[van der Hooft et al. (1996)]{vanderhooft96} 
van der Hooft F. et al., 1996, \apj, 458, L75

\bibitem[Warner(1986)]{warner86}
Warner, B. 1986, \mnras, 222, 11

\bibitem[Whitehurst \& King(1991)]{whitehurst91}
Whitehurst R., King A., 1991, \mnras, 249, 25

\bibitem[Wu et al. (2010)]{wu10} 
Wu Y.X., Yu W., Li T.P., Maccarone T.J., Li X.D., 2010, \apj, 718, 620     

\bibitem[Yanes-Rizo et al.(2022)]{yanes-rizo22}
Yanes-Rizo I.V. et al., 2022, \mnras, 517, 1476

\bibitem[Yu \& Yan (2009)]{yu09} 
Yu W., Yan Z., 2009, \apj, 701, 1940

\bibitem[Yungelson et al. (2006)]{yungelson06} 
Yungelson R.L. et al., 2006, A\&A, 454, 559

\bibitem[Zurita et al.(2002)]{zurita02}
Zurita C. et al., 2002, \mnras, 333, 791

\bibitem[Zurita et al.(2003)]{zurita03}
Zurita C., Casares J., Shahbaz T., 2003, \apj, 582, 369

\bibitem[Zurita et al.(2008)]{zurita08}
Zurita C., Durant M., Torres M.A.P., Shahbaz T., Casares J., Steeghs D., 2008, \apj, 681, 1458

\bibitem[Zurita et al.(2015)]{zurita15}
Zurita C., Corral-Santana J.M., Casares J., 2015, \mnras, 454, 3351
	
\end{thebibliography}



\appendix

\section{\ha~spectroscopy of the 2017 outburst}
\label{ap:ha}

\begin{figure}
	\includegraphics[angle=-90,width=\columnwidth]{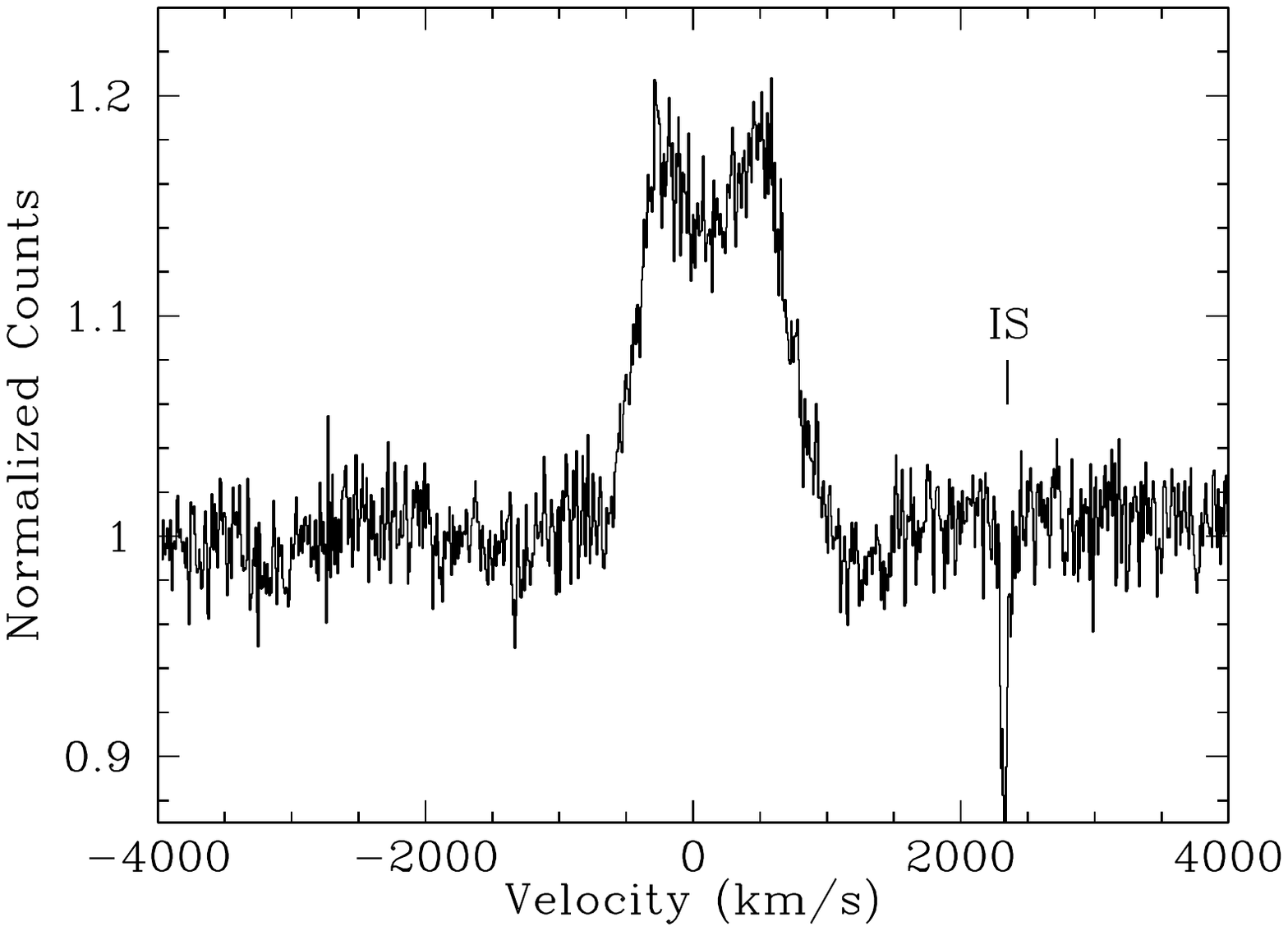}
    \caption{\ha~profile on the night of 2017 May 19. The sharp absorption to the right of \ha~corresponds to the 
    $\lambda$6613 IS band.} 
    \label{fig:a1}
\end{figure}

High resolution (46 \kms) optical spectroscopy of GRS1716 was obtained with the X-shooter spectrograph 
in the course of the 2017 outburst. One spectrum was acquired on the night of May 19 while ten others on June 3. 
We refer to \cite{cuneo20} for technical details on the acquisition and reduction of the data. 
All the spectra have been normalized to the continuum level. 
The May 19 spectrum displays a classic double-peaked profile , characteristic of an accretion 
disc (Fig.~\ref{fig:a1}). The line is moderately narrow, with FWHM=1106$\pm$9 \kms, as measured through a 
Gaussian fit. We also derive a double peak separation of 650$\pm$3 \kms~after fitting a double Gaussian model 
to the profile. 
  
\begin{figure}
	\includegraphics[angle=-90,width=\columnwidth]{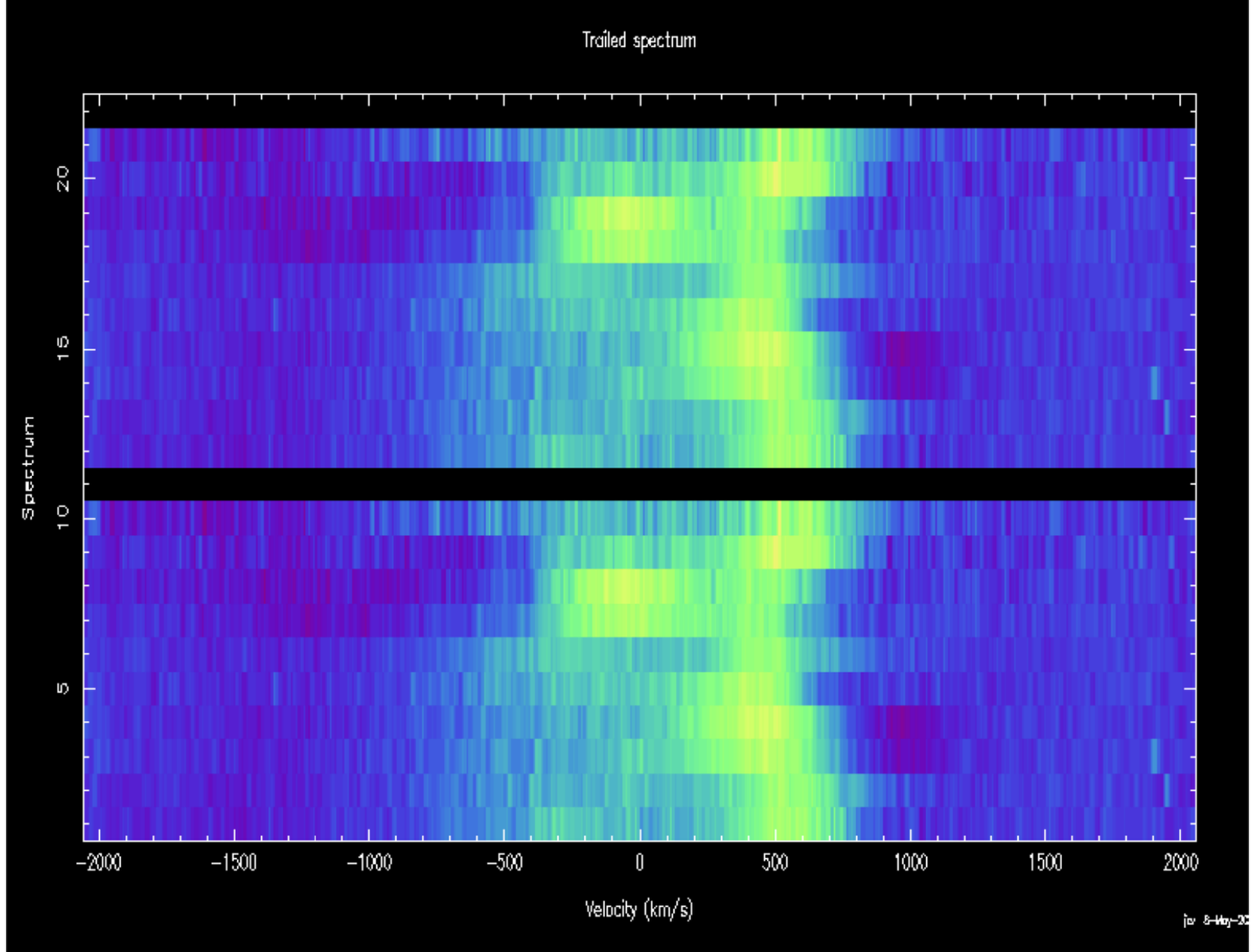}
    \caption{Trailed spectra of the \ha~line on the night of 2017 June 3. The 10 individual spectra are replicated 
    over two  orbital cycles. The line fux has been normalized by its EW to remove large contrast variations,  
   with a factor $\sim$2 amplitude. These are dominated by changes in the intensity of the red peak.} 
    \label{fig:a2}
\end{figure}

While \cite{cuneo20} focus on the detection of outflow signatures in the Balmer lines we here search for evidence  
of the orbital period by looking at variability in the radial velocities and EW of the \ha~emission, the strongest 
line in the entire spectrum. 
The ten 916 sec spectra from June 3 span over a continuous interval of $\sim$6 h and therefore nearly cover a full 
orbital cycle. Fig.~\ref{fig:a2} presents the trailed spectra of the \ha~line on that night, with two orbital 
cycles being plotted. 
We note that the blue part of the line is strongly affected by transient P-Cygni absorptions (see the skewed profiles in 
Fig.4 of \citealt{cuneo20}) although the red peak remains largely unaffected. Visual inspection shows that the latter traces a 
clear sinusoidal wave.

\begin{figure}
	\includegraphics[angle=-90,width=\columnwidth]{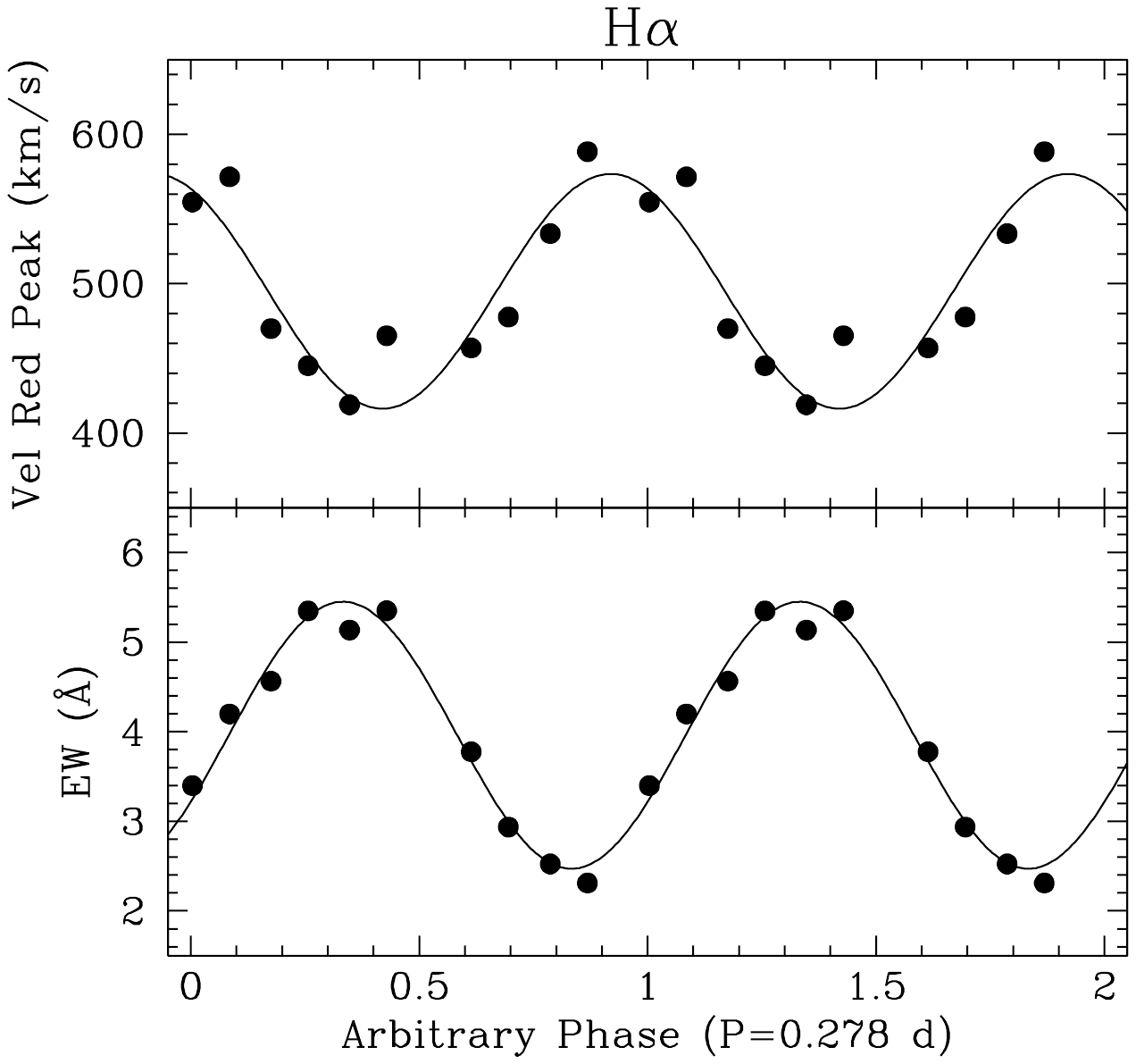}
    \caption{Top: radial velocities of the red peak in the \ha~profile. The data are folded onto two cycles, with 
    phase zero arbitrarily assigned to the time of the first spectrum. A sine wave fit with P=0.278 d is overplotted. 
    Bottom: EWs of the \ha~line folded onto two cycles, with the best sine wave fit superimposed. 
    Errorbars are always smaller than the symbol size.} 
    \label{fig:a3}
\end{figure}

In order to extract the red peak velocities we performed a two-Gaussian model fit to the individual profiles. 
The model consists of a broad base and a narrow redshifted component with all the parameters left free. 
The EW of the entire profile was also measured by integrating 
the line in a band of $\pm$1500 \kms~around the rest velocity. Fig.~\ref{fig:a3} displays the red peak velocities and 
EWs of the \ha~line, folded on the 0.278 d orbital 
period. Although our limited time coverage prevents us from 
measuring an accurate period, the observed variability is clearly consistent with 0.278 d 
and not with the 0.613 d period 
reported by M96, unless the \ha~velocity semi-amplitude were implausibly large at $\sim$235 \kms. 
Sine wave fits to the red peak velocities and EWs, with the period fixed to 0.278 d, 
gives semi-amplitudes of 79$\pm2$ kms~and 1.49$\pm0.02$  \AA, respectively. 

\section{Interstellar extinction}
\label{ap:extinction}

To constrain the reddening towards GRS1716 we have measured the strength of several IS absorption lines in the 
average of all the X-shooter spectra. For the case of the IS line at $\lambda$6613 (Fig.~\ref{fig:a1}) we find 
EW=0.20$\pm$0.01 \AA~which, according to the  linear calibration of \cite{herbig75}, corresponds to 
$E(B-V)=0.8\pm0.2$. The quoted error in EW is a conservative 
estimate based on a range of possible continuum positions, while the uncertainty in $E(B-V)$ is determined by the range of 
fitting coefficients listed in Table 4 of \cite{herbig75}. On the other hand, the Na D1 line at  
$\lambda$5896 yields EW=0.64$\pm$0.1 \AA~which translates into $E(B-V)>0.75$ \citep{munari97}. We take 
this value as a conservative lower limit because the line is most likely saturated. 
Evidence for this is provided by the ratio EW(D2)/EW(D1)=1.1. This should be $\sim$2 at low optical depths while it 
converges asymptotically towards 1.1 at large reddening values. 

The K I line at $\lambda$7699 offers a more robust diagnostic for the case of high reddening values \citep{munari97}. 
We measure EW(K I)=0.26$\pm$0.01 \AA~in our spectrum which, 
according to the empirical fit of \cite{munari97} results in $E(B-V)=1.05\pm0.05$. This value agrees well with the 
estimate of \cite{dellavalle94}, who derive $E(B-V)=0.9\pm0.2$ from a combination of different methods, including the 
strength of several IS lines, the outburst $(B-V)$ colour, the Balmer decrement and the column density of atomic hydrogen 
and molecular CO. It also agrees with the hydrogen column density N$_{\rm H}=0.70\times10^{22}$ cm$^{-2}$ measured 
during the 2017 outburst \citep{bassi19, bassi20}, which implies $E(B-V)=1.03$ \citep{guver09}. 
Considering all this information, we decide to adopt $E(B-V)=1.0\pm0.1$ as our best estimate for the interstellar 
reddening towards GRS1716.


\bsp	
\label{lastpage}
\end{document}